\documentclass[fleqn,usenatbib]{mnras}

\usepackage{newtxtext,newtxmath}

\usepackage[T1]{fontenc}
\usepackage{ae,aecompl}
\usepackage{subfigure}

\usepackage{graphicx}	
\usepackage{amsmath}	
\usepackage{amssymb}	
\usepackage{soul}
\usepackage{makecell}

\title[Star Formation in Luminous LoBALs]{Star Formation in Luminous LoBAL Quasars at 2.0 $<$ z $<$ 2.5}

\author[C. F. Wethers et al.]{
Clare F. Wethers,$^{1}$\thanks{E-mail: clare.wethers@utu.fi}
Jari Kotilainen,$^{1,2}$
Malte Schramm $^{3}$
and Andreas Schulze $^{3}$
\\
$^{1}$ Finnish Centre for Astronomy with ESO (FINCA), Vesilinnantie 5, FI-20014 University of Turku, Finland \\
$^{2}$ Department of Physics and Astronomy, Vesilinnantie 5, FI-20014 University of Turku, Finland \\
$^{3}$ National Astronomical Observatory of Japan, Mitaka, Tokyo 181-8588, Japan \\
}

\date{Accepted XXX. Received YYY; in original form ZZZ}

\pubyear{2019}

\begin{document}
\label{firstpage}
\pagerange{\pageref{firstpage}--\pageref{lastpage}}
\maketitle

\begin{abstract}
Low-ionisation broad absorption line quasars (LoBALs) mark an important, yet poorly understood, population of quasars showing direct evidence for energetic mass outflows. 
We outline a sample of 12 luminous (L$_{\rm{bol}}$ $>$10$^{46}$ergs$^{-1}$) LoBALs at 2.0$<$z$<$2.5 - a key epoch in both star formation and black hole accretion, which have been imaged as part of a targeted program with the \textit{Herschel} Spectral and Photometric Imaging REceiver (SPIRE). 
We present K-band NOTCam spectra for three of these targets, calculating their spectroscopic redshifts, black hole masses and bolometric luminosities, and increasing the total number of LoBAL targets in our sample with spectral information from five to eight. 
Based on FIR obeservations from \textit{Herschel} SPIRE, we derive prolific SFRs ranging 740 - 2380M$_{\rm{\odot}}$yr$^{-1}$ for the detected targets, consistent with LoBALs existing in an evolutionary phase associated with starburst activity.
Furthermore, an upper limit of $<$440M$_{\rm{\odot}}$yr$^{-1}$ is derived for the non-detections, meaning moderate-to-high SFRs cannot be ruled out, even among the undetected targets. Indeed, we detect an enhancement in both the SFRs and FIR fluxes of LoBALs compared to HiBAL and non-BAL quasars, further supporting the evolutionary LoBAL paradigm. Despite this enhancement in SFR however, the environments of LoBALs appear entirely consistent with the general galaxy population at 2.0$<$z$<$2.5.
\end{abstract}

\begin{keywords}
quasars:general -- galaxies:evolution -- galaxies: star formation -- quasars:absorption lines -- galaxies:active
\end{keywords}


\section{Introduction}
\label{sec:Intro}

The co-evolution of quasars and their host galaxies is now widely accepted, due primarily to the tight correlations observed between the mass of the super-massive black hole (M$_{\rm{BH}}$) and various properties of the galaxy bulge \citep[e.g.][]{magorrian98,ferrarese00,gebhardt00,kormendy13,graham14}. However, the scale over which quasars and their hosts appear coupled to one another extends well beyond the sphere of influence of the black hole and to date the processes by which the quasar seemingly influences the galaxy on these scales remain poorly understood. One proposed mechanism is the presence of quasar outflows: energetic mass outflows potentially responsible for both quenching star formation in the galaxy and self-regulating black hole growth \citep[e.g.][]{silk98,dimatteo05,fabian12,king15}. Direct observations of galaxies hosting these outflows however, remain sparse, particularly at z$\sim$2 where both black hole accretion and star formation peak \citep[e.g.][]{aird15, madau14}.

Broad absorption line quasars (BALs) are identified via their blueshifted absorption features, primarily consisting of high-ionisation lines such as C$_{\rm{IV}}$ and Si$_{\rm{IV}}$ and are thought to exist in $\sim$15 per cent of optically selected quasars \citep[e.g.][]{hewett03,gibson09}, although the true fraction may be as high as $\sim$40 per cent \cite{allen11}. Generally, BALs are classified into two categories - high-ionisation BALs (HiBALs), containing only high-ionisation absorption features in their spectra and accounting for $\sim$85 per cent of the total BAL population, and low-ionisation BALs (LoBALs), whose spectra additionally contain broad absorption features from low-ionisation ions such as Mg$_{\rm{II}}$ and Al$_{\rm{III}}$. BALs represent an important yet poorly understood population of quasars showing direct evidence for energetic mass outflows launched as radiation-driven disc winds \citep{proga00,proga04}.

The role these outflows play in galaxy evolution remains a key question in quasar-galaxy co-evolution. On the one hand, LoBAL winds may be responsible for quenching star formation in the galaxy, blowing out gas and dust from the galaxy and thus restricting the material available to form stars. Indeed, \cite{farrah12} find an anti-correlation between the strength of the quasar outflow and star formation in its host, concluding the strongest outflows to reside in the most quiescent galaxies. In semi-analytic models of galaxy assembly however \citep{granato04}, these outflows are invoked not only to remove dense gas from the galaxy centre, but also to provide metal enrichment to the intergalactic medium (IGM). Quasar outflows may therefore also work to trigger regions of star formation in the galaxy by compressing cool, metal rich gas and allowing stars to form. In many cases however, it has been shown that the the IGM is unable to cool efficiently and therefore cannot fall back into the galaxy to fuel star formation in this manner \citep{gabel06}. Furthermore, simulations by \cite{zubovas17} find that sufficiently luminous outflows are capable of shutting off fragmentation whilst the quasar is active, thus restricting star formation in the galaxy. Only once the quasar has `shut off' is gas compression shown to result in a burst of star forming activity. During the active quasar phase, we therefore expect outflows to predominantly work to quench star formation in the galaxy.

Two main interpretations of the BAL phenomenon exist. The first is an orientation scenario, whereby BAL winds are thought to exist in  all quasars, but can only be viewed along particular sight lines due to the low covering factor of the quasar's broad absorption line region (BALR) \citep[e.g.][]{voit94,becker00}. This scenario is consistent with a unified model of quasars \citep{antonucci93} and appears to provide a good model for the HiBAL population. In particular, this model explains the strong similarities observed between HiBAL and non-BAL spectra \citep[e.g.][]{weymann91,reichard03} and the lack of enhancement in the millimetre detection rates of HiBALs \citep[e.g.][]{priddey07,willott03,lewis03}. On the other hand, BALs have been observed at a wide range of inclinations \citep{ogle99,dipompeo10,dipompeo11}, directly contradicting a key prediction of the orientation scenario. A second, alternative interpretation suggests that BALs, particularly LoBALs, mark a distinct phase in the early lifetime of a quasar, existing in a short-lived transition period between a merger-induced starburst galaxy and an UV-luminous quasar \citep[e.g.][]{boroson92}. 

One way in which the two paradigms - orientation and evolution - may be distinguished is through their FIR fluxes. If LoBALs are indeed being observed following an epoch of enhanced star formation, we would expect to yield higher average FIR flux densities compared to non-BAL quasars due to the high dust masses within the LoBAL host galaxy. As such, FIR flux densities may be used to distinguish between orientation and evolution scenarios \citep{cao12}. Whilst several high-redshift (z$>$4) studies \citep[e.g.][]{omont96,carilli01} find tentative evidence suggesting BALs to be more luminous at millimetre wavelengths compared to non-BALs, others find no differences in the mid-IR fluxes of BALs and non-BALs \citep{gallagher07} based on Spitzer observations. We note however, that previous studies have predominantly focused on populations of HiBALs, with relatively few considering LoBALs at z$\sim$2.0. In terms of LoBAL properties, an increase in FIR flux density is likely to translate into an enhancement in their star formation, as high dust masses are often associated with star formation in the galaxy. Indeed, enhanced star formation tracing the decaying starburst within populations of dust obscured quasars is a key prediction of the evolutionary model \citep[e.g.][]{farrah07,lipari94}. \cite{canalizo01}, for example, find evidence for strong recent star formation in LoBALs at z$<$0.4, which appears to be directly linked to tidal interactions in the galaxy. At higher redshifts however, \cite{schulze17} find no statistical differences in the distributions of either M$_{\rm{BH}}$ or Eddington ratios of LoBAL quasars at z$\sim$2.0 compared to a matched sample of non-BAL quasars, implying that LoBALs do not comprise a distinct population but rather may exist as part of the orientation paradigm. It remains to be seen however, whether these z$\sim$2.0 LoBALs exhibit a similar enhancement in star formation to their low redshift counterparts \citep{canalizo01}. To this end, we estimate the FIR SFRs for a sample of 12 LoBALs at 2.0 $<$ z $<$ 2.5. In particular, we search for evidence of enhanced star formation in these galaxies compared to non-BAL and HiBAL populations, which may indicate starburst or post-starburst activity consistent with an evolutionary picture of LoBALs.

This paper is structured as follows. Sec.~\ref{sec:Data} details the LoBAL sample considered in this work. Results are presented in Sec.~\ref{sec:Results} and discussed in Sec.~\ref{sec:Discussion}, where our findings are directly compared to independent studies of HiBAL and non-BAL quasars to test the evolutionary interpretation of LoBALs. Our key findings are summarised in Sec.~\ref{sec:Conclusions}. Throughout this paper, we assume a flat $\Lambda$CDM cosmology with $H_{0}$ = 70 km s$^{-1}$ Mpc$^{-1}$, $\Omega_{M}$ = 0.3 and $\Omega_{\Lambda}$ = 0.7. Unless otherwise specified, all quoted magnitudes are given in the AB system.

\section{Data}
\label{sec:Data}

This work concerns a sample of 12 LoBALs, drawn from the BAL quasar catalogue of \cite{allen11}. The initial selection criteria is outlined in detail in the work of \cite{schulze17}, but is summarised below.

\subsection{Sample Selection}
\label{sec:SampleSelect}

LoBALs were initially identified from their SDSS DR6 optical spectra \citep{schneider07,schneider10}, requiring a non-zero balnicity index (BI) as defined by \cite{weymann91}, such that BI$>$0 in either Mg$_{\rm{II}}$ ($\lambda$2800) or Al$_{\rm{III}}$ ($\lambda$1860). Based on this definition, \cite{allen11} identify 368 LoBALs, of which 24 lie in the redshift window 2.0 $<$ z $<$ 2.5, marking a key epoch in both star formation and black hole accretion \citep{aird15,madau14}. From these, we select a sub-sample of 12 LoBALs overlapping a targeted program with the \textit{Herschel} SPIRE \citep{proposal07}, in which targets were imaged deeper than the nominal 5$\sigma$ SPIRE depths of 45, 62 and 53mJy at 250, 350 and 500$\mu$m respectively.LoBALs included in this targeted program were imaged for 583s per band in the "small scan" mode (see Sec.~\ref{sec:Herschel} for details), compared to the standard 169s. Details of the 12 targets comprising the full sample of LoBALs considered in this paper, are given in Tab.~\ref{tab:DataTab}. Seven of the 12 LoBALs in our sample overlap with the LoBAL sample presented in \cite{schulze17}, where targets were further required to be detected in the $K$-band with 2MASS and to lie at z$>$2.2, such that H$\alpha$ emission line lies within the wavelength range of NOTCam (1950 $<$ $\lambda_{\rm{rest}}$ $,$ 2370\AA). Of these seven overlapping LoBALs, five have previously been observed with NOTCam on the Nordic Optical Telescope (NOT), providing supplementary IR spectra for these targets and thus information about their M$_{\rm{BH}}$ and bolometric quasar luminosity, L$_{\rm{bol}}$. For this work, we obtain similar observations for a further three targets in our sample (see Sec.~\ref{sec:NOTCam}), providing M$_{\rm{BH}}$ and  L$_{\rm{bol}}$ information for a total of eight LoBALs in our sample.

\subsection{\textit{Herschel} Data}
\label{sec:Herschel}

Throughout this paper we make use of archival \textit{Herschel} SPIRE data taken as part of a targeted quasar survey proposed by \cite{proposal07}. The SPIRE instrument operates in three bands centred at 250$\mu$m, 350$\mu$m and 500$\mu$m with a 4$\times$8 arcmin field of view \citep{pilbratt10}. All LoBAL targets considered in this paper were observed between 2010 June and 2011 January using the "small scan map" observing mode, which fully samples maps over a $<$5 arcmin diameter area with a fixed scan speed of 30 arcsecs$^{-1}$. The "small scan map" mode is identical in sensitivity to the "large scan map" mode, but its smaller area coverage makes it suitable for observing individual targets across several different fields, as is the case for our sample. This "small scan" configuration creates science-quality maps of diameter $>$5 arcsec, where the scan legs overlap. The resulting images have pixel scales 6, 10 and 14 arcsec at 250, 350 and 500$\mu$m respectively. Likewise, the SPIRE beam size varies across each band, measuring 18.1, 24.9, 36.4 arcsec at 250, 350 and 500$\mu$m respectively. Each LoBAL target in the sample has an exposure time of 583s per band. 

In Section ~\ref{sec:SFRs} we additionally make use of data from the \textit{Herschel} photodetector Array Camera and Spectrometer PACS instrument at 70$\mu$m (blue band) and 160$\mu$m (red band), tracing the emission from warm dust at the redshift of our LoBAL sample \citep{pilbratt10}. PACS observations were taken as part of the same targeted quasar survey proposed by \cite{proposal07} between 2010 June and November. All observations were were taken in the `PACSPhoto' scan map mode with a medium scan speed of 20 arcsecs$^{-1}$ and a scan angle of 70 degrees, with an angle of 110 degrees for the cross-scan. The resulting observations cover a 1.75 $\times$ 3.5 arcmin field of view in each band, with pixel scales of 3.2 and 6.4 arcsec for the blue and red bands respectively. Each band has a total exposure time of 276s, reaching a typical 5$\sigma$ flux sensitivity of $\sim$5mJy at 70$\mu$m and $\sim$10mJy at 160$\mu$m. 

\begin{table*}
	\centering
	\caption{Overview of the \textit{Herschel} SPIRE observations for the sample of 12 LoBALs. For the eight targets for which NOTCam spectra are available, the quoted redshifts are spectroscopically determined from the H$\alpha$ line emission (\protect\cite{schulze17} and this work). Redshifts for the remaining four targets are instead taken from the photometric estimates of \protect\cite{allen11}. Aperture photometry for each of the LoBAL targets (mJy) assumes aperture radii of 22, 30 and 40 arcsec for the 250, 350 and 500$\mu$m images respectively, centred on the catalogued position of the source. Quoted uncertainties are derived from the deviation in the integrated flux of randomly placed apertures, lying within 1.5 arcsec of the source. `*' marks targets detected at $>$5$\sigma$ at the specified wavelength.}
	\label{tab:DataTab}
	\begin{tabular}{lrrrrrrrcrrr}
        \thead{Name} & \thead{RA} & \thead{Dec} & \thead{Obs. Date} & \thead{S$_{250}$ $\pm$ $\sigma_{250}$ \\ {[mJy]}} & \thead{S$_{350}$ $\pm$ $\sigma_{350}$ \\ {[mJy]}} & \thead{S$_{500}$ $\pm$ $\sigma_{500}$ \\  {[mJy]}} & z & \thead{log(M$_{\rm{BH}}$) \\ {[M$_\odot$]}} & \thead{log(L$_{\rm{bol}}$) \\ {[ergs$^{-1}$]}} & \thead{log$\lambda$} \\
		\hline
        SDSSJ0753+2102 & 118.2935 & 21.0457 & 2010.09.21  & 32.93  $^{+ 8.62}_{- 8.27}$  & * 41.40 $^{+ 6.86}_{- 8.88}$  & 28.79  $^{+ 1.22}_{- 4.20}$  & 2.290 & 9.43 & 46.23 & -1.30 \\
        SDSSJ0810+4806 & 122.6031 & 48.1043 & 2010.09.21  & * 30.98  $^{+ 7.90}_{- 6.21}$  &  * 44.54 $^{+14.84}_{-10.31}$ & * 41.71  $^{+21.35}_{-27.50}$   & 2.256 & - & - & - \\
        SDSSJ0839+0454 & 129.8567 & 4.9056  & 2010.10.11  & * 52.39  $^{+ 9.69}_{- 8.20}$  &  * 77.42 $^{+29.50}_{-12.62}$ & * 60.59  $^{+26.00}_{- 9.94}$   & 2.447 & - & - & - \\
        SDSSJ0943-0100 & 145.9092 & -1.0054 & 2010.11.09  & * 87.22  $^{+10.74}_{-14.32}$  &  * 94.46 $^{+ 8.52}_{-17.62}$ & * 56.53  $^{+15.26}_{- 8.82}$   & 2.376 & 9.55 & 47.47 & -0.18\\
        SDSSJ0957+5120& 149.3026 & 51.3497 & 2010.10.11   & -1.80  $^{+ 9.45}_{- 6.02}$  &  21.54 $^{+11.20}_{-13.12}$ & -15.61 $^{+ 7.28}_{-19.33}$   & 2.116 & - & - & - \\
        SDSSJ1011+5155 & 152.7871 & 51.9316 & 2010.10.11  & 19.64  $^{+7.03}_{-5.79}$  &  26.26 $^{+10.20}_{-6.37}$ & 21.05  $^{+ 6.87}_{- 3.04}$   & 2.465 & 9.71 & 47.16 & -0.66 \\	
        SDSSJ10285+5110 & 157.2097 & 51.1814 & 2010.06.13 & 12.69  $^{+ 9.86}_{- 9.44}$  &  22.75 $^{+ 7.74}_{- 5.48}$ & 19.26  $^{+2.85}_{-13.70}$   & 2.426 & 9.38 & 47.50 & 0.02 \\	
        SDSSJ1132+0104 & 173.0538 &  1.0782 & 2010.06.28  & * 29.27  $^{+10.02}_{- 6.41}$  &  * 34.72 $^{+ 9.31}_{- 6.84}$ & 24.06  $^{+12.19}_{- 8.94}$   & 2.328 & 9.63 & 47.27 & -0.46 \\	
        SDSSJ1341-0036 & 205.4380 & -0.6087 & 2010.07.26  & 12.58  $^{+ 9.66}_{- 7.13}$  &  35.42 $^{+ 9.58}_{-10.75}$ & 26.00  $^{+ 8.36}_{-10.82}$   & 2.215 & - & - & - \\	
        SDSSJ1352+4239 & 208.1932 & 42.6566 & 2010.06.29  & 41.11  $^{+11.09}_{- 3.67}$  &  34.46 $^{+ 8.25}_{- 5.22}$ & 7.92   $^{+ 9.61}_{- 4.55}$   & 2.261 & 9.53  & 47.07 & -0.55 \\
        SDSSJ1516+0029 & 229.1533 &  0.4946 & 2011.01.30  & 3.33   $^{+ 8.80}_{- 9.84}$  &  25.54 $^{+ 6.62}_{- 8.96}$ & 22.77  $^{+16.52}_{- 5.73}$   & 2.251 & 9.49 & 47.37 & -0.22 \\	
        SDSSJ1723+5553 & 260.9212 & 55.8946 & 2010.06.29  & -11.36 $^{+ 4.31}_{- 4.56}$  &  -2.28 $^{+ 6.71}_{- 6.35}$ & -9.53  $^{+ 1.60}_{- 4.89}$   & 2.108 & 9.20 & 47.18 & -0.12 \\
		\hline
	\end{tabular}
\end{table*}
 
\subsection{Supplementary NOTCam Data}
\label{sec:NOTCam}

Already, NIR spectra exist for five of the 12 targets in our LoBAL sample \citep{schulze17}. Using NOTCam on the NOT, we obtain low-resolution (R=2500) K-band spectroscopy for an additional three targets: SDSSJ075310+210244, SDSSJ135246+423923 and SDSSJ172341+555340, increasing the number of LoBAL targets in our sample with NIR spectra from five \citep{schulze17} to eight. Observations were carried out in service mode over two nights from 2019 April 10th to 2019 April 12th under good seeing conditions ($<$1.0 arcsec), using a 0.6 arcsec (128$\mu$m wide-field) slit. Total exposure times range from 78 minutes for the brightest target to 130 minutes. Telluric standards lying at a similar airmass and sky position were observed either directly before or after each of the quasar observations and an ABABAB (AB3) dither pattern was performed along the slit for each target to improve the sky subtraction. Data reduction was performed following the standard reduction steps for sky subtraction, flat-fielding and telluric correction using the relevant IRAF software. Following the wavelength calibration, based on either an Ar or Xe arc lamp, the 1D spectrum for each target was extracted (Fig.~\ref{fig:spec}).

\section{Results}
\label{sec:Results}

\subsection{Black Hole Masses}
\label{sec:Mbh}

\begin{figure}
    \centering
    \includegraphics[trim= 10 10 20 30 ,clip,width=.5\textwidth]{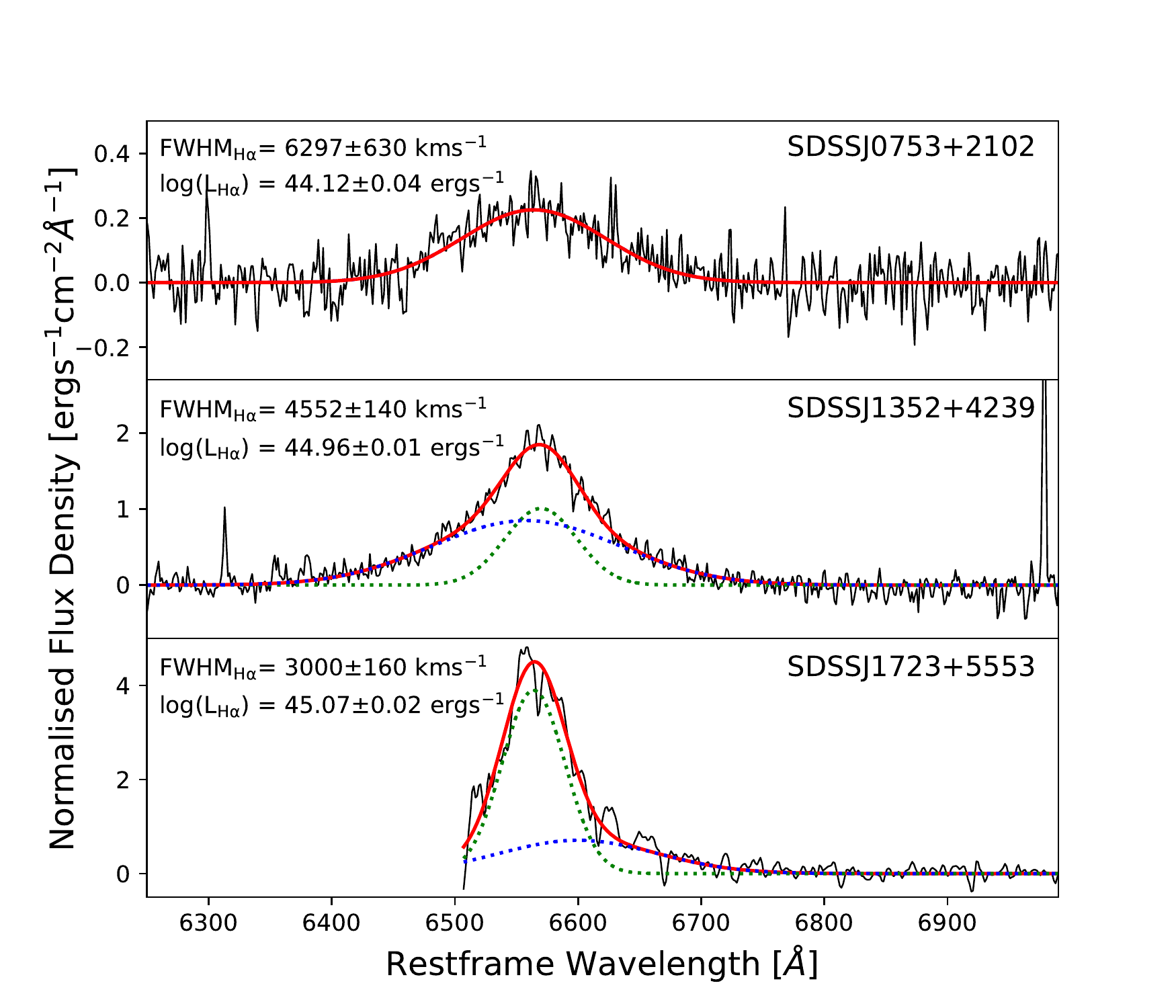}
    \caption{1D NOTCam 6000-7000\AA\ continuum-subtracted spectra \textit{(black)} for SDSSJ075310+210244 \textit{(upper)}, SDSSJ135246+423923 \textit{(middle)} and SDSSJ172341+555340 \textit{(lower)}. The broad Gaussian components used in the fit are also shown (\textit{dotted lines}).}
    \label{fig:spec}
\end{figure}

Fig.~\ref{fig:spec} shows the extracted continuum-subtracted spectra for each of our quasar targets. In order to estimate the H$\alpha$ line width, we begin by fitting a local power law relation to the continuum of the NOTCam spectra and remove this contribution from the spectra. The H$\alpha$ emission feature is then fitted by a set of broad (>1800 kms$^{-1}$) Gaussian components and the full-width half-maximum (FWHM) is measured from the profile of the combined Gaussians. No NII component is included in the fitting, as its contribution is found to be negligible. We opt to use the FWHM as its dependency on the wings of the emission profile is much weaker than other methods, and thus provides a more robust estimate of the line width at low S/N. The luminosity of the H$\alpha$ line emission, L$_{\rm{H\alpha}}$, is derived directly from the strength of the H$\alpha$ emission feature and M$_{\rm{BH}}$ estimates are based on the virial method for single-epoch broad-line AGN spectra, following the methods outlined in \cite{schulze17}. M$_{\rm{BH}}$ estimates are dependent on both the width and luminosity of the broad H$\alpha$ feature, such that

\begin{equation}
M_{\rm{BH}} = 10^{6.711} \left( \frac{L_{\rm{H\alpha}}}{10^{42} \rm{ergs^{-1}}} \right) ^{0.48} \left( \frac{\rm{FWHM}_{\rm{H\alpha}}}{1000 \rm{kms^{-1}}} \right) ^{2.12} M_{\rm{\odot}}.
\end{equation}

The bolometric quasar luminosity, $L_{\rm{bol}}$, is then derived from $L_{\rm{H\alpha}}$ adopting a bolometric correction factor such that $L_{\rm{bol}}$ = 130$L_{\rm{H\alpha}}$ \citep[e.g.][]{stern12}. The resulting estimates for $M_{\rm{BH}}$ and $L_{\rm{bol}}$ are presented for SDSSJ075310+210244, SDSSJ135246+423923 and SDSSJ172341+555340 in Tab.~\ref{tab:DataTab}, along with those previously derived by \cite{schulze17}. We additionally calculate the Eddington ratio ($\lambda$) for each of the eight targets for which M$_{\rm{BH}}$ are obtained, finding these values to be consistent with those presented for the z$\sim$ 2.3 sample in \cite{schulze17}. Across our sample, we calculate Eddington ratios ranging 0.05$<$ $\lambda$ $<$ 1.05, with a mean value of 0.50. By comparison, \cite{schulze17} derive values ranging 0.22$<$ $\lambda$ $<$ 1.02, with a mean value of 0.51.

\subsection{Aperture Photometry}
\label{sec:AperPhot}

\textit{Herschel} SPIRE 250$\mu$m, 350$\mu$m and 500$\mu$m observations are downloaded from the \textit{Herschel} Science Archive (HSA) and processed using the \textit{Herschel} Interactive Processing Environment (HIPE). The background flux in each band is estimated from a blank area of sky close to the target, selected by eye, and is subtracted from each pixel in the image. Aperture photometry is performed on these background-subtracted images in the HIPE environment, summing the flux within circular apertures of radius 22, 30 and 40 arcsec at 250$\mu$m, 350$\mu$m and 500$\mu$m respectively, in accordance with the methods outlined in \cite{pearson14}. Flux corrections are applied to account for the shape of the beam, K$_{\rm{beam}}$, and the colour of the SED, K$_{\rm{col}}$, the values for which are detailed in the \textit{Herschel} SPIRE handbook\footnote{http://herschel.esac.esa.int/Docs/SPIRE/html/spire\_om.html}. No aperture correction, K$_{\rm{aper}}$, is required, as the extent of our z$\sim$2.2 galaxies lie well within the aperture radii. To estimate any photometric uncertainties, circular apertures matching those centred on the target are randomly positioned on the sky. Fluxes are measured within each of these sky apertures and the 16th and 84th percentiles are taken as the 1$\sigma$ lower and upper uncertainties on the photometry respectively, accounting for the skewed flux distribution and thus minimising the contribution of any spurious sources in the image. In order to accurately represent the error within the central region of the image, sky apertures are placed close to the target ($<$1.5 arcmin), whilst avoiding regions which overlap the photometric aperture. The resulting aperture photometry for the LoBAL sample is given in Table~\ref{tab:DataTab}, along with the associated uncertainties. 

\begin{figure}
	\centering 
	
	\subfigure{\includegraphics[trim= 125 0 110 0  ,clip,width=.5\textwidth]{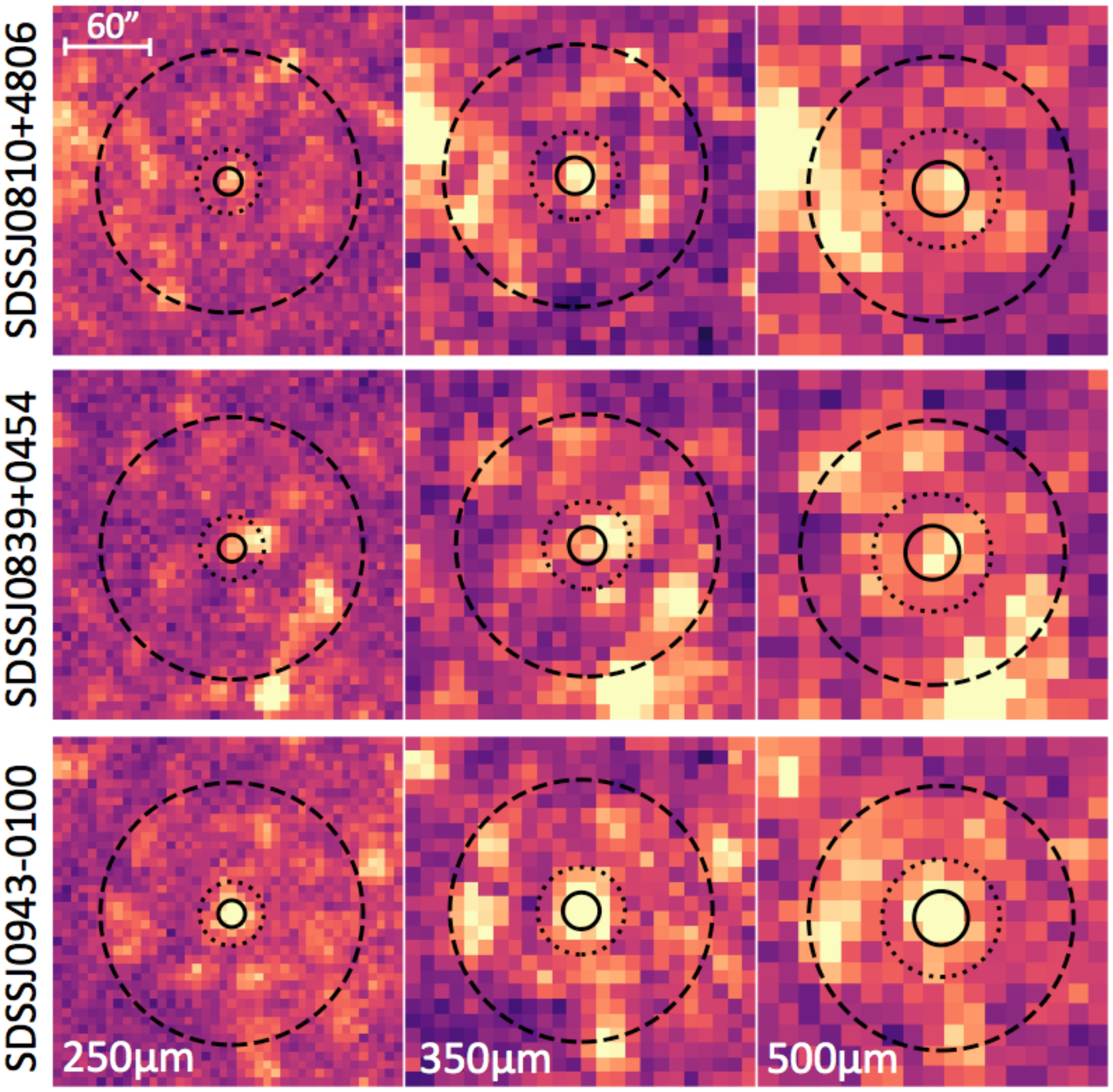}}
	
	\subfigure{\includegraphics[trim= 30 290 30 215 ,clip,width=.5\textwidth]{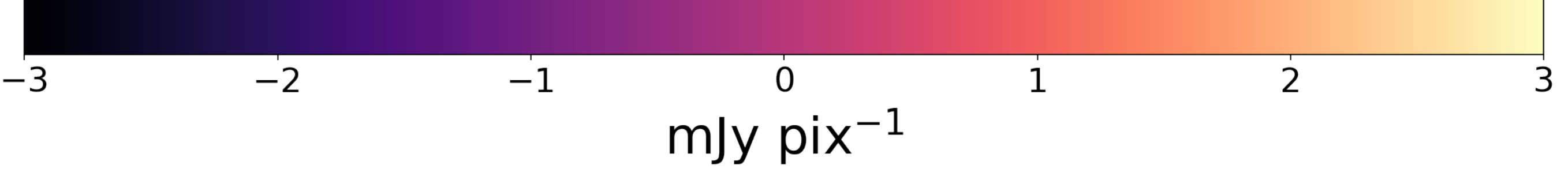}}
	
\caption{The three LoBALs detected at $>$5$\sigma$ in all Herschel SPIRE bands (250, 350 and 500$\mu$m). Solid circle shows size and position of Herschel SPIRE beam in each band (of area 250$\mu$m: 465.39, 350$\mu$m: 822.58, 500$\mu$m: 1768.66 arcsec$^2$). Dotted circle denotes aperture size used to derive the photometry (of radius 250$\mu$m: 22, 350$\mu$m: 30, 500$\mu$m: 40 arcsec). Dashed circles denote the 1.5 arcmin radius areas in which apertures were placed to estimate the uncertainty on the derived source fluxes. North is up, East is left.}

\label{fig:ImgDetect}
\end{figure}

To determine which of the 12 LoBAL sources are detected with \textit{Herschel} SPIRE, signal-to-noise (S/N) maps are created from the HIPE-processed images. A $>$5$\sigma$ detection threshold is selected, such that no negative sources are detected with the same confidence in any of the images. Of the 12 LoBAL targets considered, three are detected above this threshold in all three SPIRE bands. These three targets - SDSSJ0810+4806, SDSSJ0839+0454 and SDSSJ0943-0100 - are therefore classed as detections and form a core subsample of this work (Fig.~\ref{fig:ImgDetect}). We do however note that the flux of SDSSJ0839+0454 appears to be dominated by light from the source North West of the catalogued quasar position. To confirm that this flux is associated with the quasar, we search for SDSSJ0839+0454 observations from the Wide-field Infrared Survey Explorer (WISE). Indeed, we find our quasar target is detected in three of the four WISE bands (W1;$\lambda$3.4$\mu$m, W2;$\lambda$4.6$\mu$m, W3;$\lambda$12.0$\mu$m) and we detect positive flux at the position of our target in the remaining band (W4;$\lambda$22.0$\mu$m). In contrast, no source is detected in multiple WISE bands north-west of the quasar target. We therefore conclude that the although the flux in Fig~\ref{fig:ImgDetect} appears slightly offset from the catalogue position of SDSSJ0839+0454, it is still likely associated with the quasar and not a spurious source in the image. In addition to the three LoBALs detected across all SPIRE bands, one target - SDSSJ1132+0104 - is detected at both 250 and 350$\mu$m and another - SDSSJ0753+2102 - at only 350$\mu$m. All detections in each band are marked with an `*' in Table~\ref{tab:DataTab}. Later in Sec.~\ref{sec:caveats}, we shall discuss the potential effects of source blending to our results, but for now we assume the measured aperture fluxes arise solely from the LoBAL target.

\subsection{Image Stacking}
\label{sec:ImgStack}

\begin{figure}
	\centering 
	\subfigure{\includegraphics[trim= 80 23 40 25,clip,width=.5\textwidth]{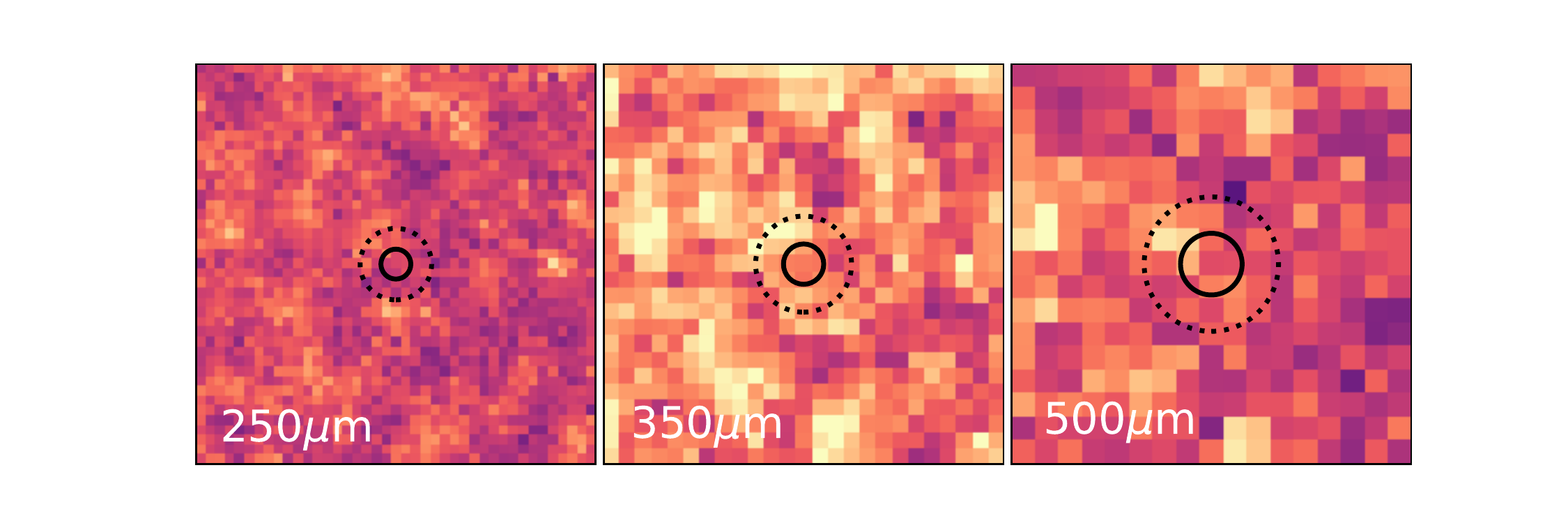}}
	
	\subfigure{\includegraphics[trim= 45 100 10 0,clip,width=.5\textwidth]{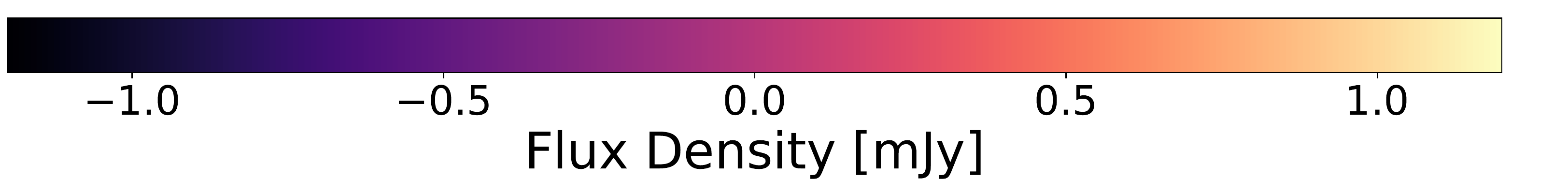}}
	
\caption{Inverse-varience mean-weighted stack of the seven non-detected LoBAL targets. Bright sources near to each target have been masked in order to prevent spurious sources in the stacked image. Beam (\textit{solid circle}) and aperture (\textit{dotted circle}) sizes in each band are overlaid for reference. North is up, East is left.}

\label{fig:ImgStack}
\end{figure}

Seven of the 12 targets in our LoBAL sample (58 per cent) remain undetected across all \textit{Herschel} SPIRE bands. In order to explore the average properties of our LoBAL sample, we therefore stack these non-detections, convolving each of the HIPE-processed images with a normalised PSF model\footnote{PSFs for each band are taken from http://herschel.esac.esa.int /twiki/bin/view/Public/SpirePhotometerBeamProfileAnalysis2`}. A background subtraction is performed for each individual target prior to stacking, following the methods outlined in Sec.~\ref{sec:AperPhot}. In the case of SDSSJ1352+4239, we also note a bright spurious source located North East of the central target, which we mask prior to stacking to ensure it does not impact the inferred stacked flux. The masked, background-subtracted images are then co-added such that each pixel in the stacked image denotes the mean-weighted flux density of that pixel in a given band. An inverse variance weighting is applied to the stack to account for any noise variation across the sample i.e. 

\begin{equation}
S_{\rm{ij}} = \frac{\sum_{\rm{k=1}}^{\rm{N}} P_{\rm{ij}}^{\rm{k}} \times f_{\rm{ij}}^{\rm{k}} /  (\sigma_{\rm{ij}}^{\rm{k}} \times \sigma_{\rm{ij}}^{\rm{k}})}{\sum_{\rm{k=1}}^{\rm{N}} P_{\rm{i,j}}^{\rm{k}} /  (\sigma_{\rm{ij}}^{\rm{k}} \times \sigma_{\rm{ij}}^{\rm{k}})},
\end{equation}

\noindent where $S_{\rm{ij}}$ denotes the stacked flux at the centre of the image. $P_{ij}$ denotes the response function of \textit{Herschel}, estimated as the PSF in each band. $f_{ij}$ are the flux densities for each of the $N$ sources in our sample and $\sigma_{ij}$ are the corresponding noise maps for each target in the stack. The stacked non-detections are presented in Fig.~\ref{fig:ImgStack}, from which the mean-weighted flux density is measured from the central pixel of the stacked image, corresponding to the pixel closest to the catalog position of the target. We derive mean-weighted flux densities of 0.21$\pm$0.21, 0.54$\pm$0.33 and 0.26$\pm$0.34 mJy/pix at 250, 350 and 500$\mu$m respectively, where the quoted uncertainties denote the standard deviation of pixels within a blank patch of sky in the stacked image, prior to applying the PSF weighting. Apertures are centred on the catalogue position of each target to measure the total integrated flux in each band, following the methods outlined in Sec.~\ref{sec:AperPhot}. This returns aperture fluxes of 10.74$^{+1.96}_{-2.09}$, 20.65$^{+1.39}_{-2.45}$ and 9.32$^{+1.70}_{-2.51}$mJy for the stacked non-detections at 250, 350 and 500$\mu$m respectively.

\subsection{FIR SFRs}
\label{sec:SFRs}

If LoBALs mark an evolutionary phase associated with merger-induced starburst activity, we expect to observe an enhancement in the FIR fluxes, and thus FIR SFRs, of LoBALs compared to populations of non-BAL quasars \citep{cao12}. To this end, we estimate the FIR SFRs of our detected targets (SDSSJ0818+4806, SDSSJ0839+0454 and SDSSJ0943-0100) based on the aperture photometry derived in Sec.~\ref{sec:AperPhot} (Tab.~\ref{tab:DataTab}). At the redshift of our sample (2.0$<$z$<$2.5), the \textit{Herschel} SPIRE photometry covers restframe wavelengths 75$\lesssim$ $\lambda$ $\lesssim$ 150 $\mu$m, tracing the peak of thermal emission from star formation. Although the hotter thermal emission from the quasar is thought to rapidly drop off at these wavelengths, some studies suggest that emission from hot dust in the torus may still contribute significantly to the \textit{Herschel} SPIRE bands, particularly among populations of bright quasars. \cite{symeonidis16} for example, find the contribution from quasar heating may be comparable to that from star formation out to $\lambda$ $<$ 1000$\mu$m in the most luminous quasars. To this end, we combine the \textit{Herschel} SPIRE photometry (Tab.~\ref{tab:DataTab}) with NIR photometry from both the WISE and \textit{Herschel} PACS in order to estimate this potential quasar contamination and isolate the thermal emission from star formation in the host. We opt to use the unWISE photometry \citep{lang14}, which is derived from the de-blurred coadd images whilst preserving the original resolution of WISE, making use of this photometry in the W1 (3.4$\mu$m), W2 (4.6$\mu$m), W3 (12.0$\mu$m) and W4 (22.0$\mu$m) bands. The PACS photometry is derived in the blue (70$\mu$m) and red (160$\mu$m) bands from the UNIMAP images using the aperture photometry task within the HIPE environment. Apertures of 12 and 22 arcsec are assumed for the blue and red bands respectively, with the background estimation taken from an annulus spanning 35-45 arcsec from the catalogued source position. Uncertainties on the PACS photometry are derived following the methods outlined in Sec.~\ref{sec:AperPhot}, measuring the variation within randomly placed apertures in the PACS images. The derived PACS aperture photometry is presented in Tab.~\ref{tab:PACSphot}, along with the unWISE photometry from \cite{lang14}.

\begin{table*}
	\centering
	\caption{Details of the unWISE and PACS photometry (both in mJy) used in the SED fitting for the three LoBALs in our sample detected at $>$5$\sigma$ in all three \textit{Herschel} SPIRE bands.}
	\label{tab:PACSphot}
	\begin{tabular}{lrrrrrr}
		\hline
        Name & S$_{3.4}$ $\pm$ $\sigma_{3.4}$ & S$_{4.6}$ $\pm$ $\sigma_{4.6}$ & S$_{12.0}$$\pm$ $\sigma_{12.0}$  & S$_{22.0}$$\pm$ $\sigma_{22.0}$  & S$_{70}$$\pm$ $\sigma_{70}$  & S$_{160}$$\pm$ $\sigma_{160}$  \\
		\hline

        SDSSJ0810+4806 & 0.21 $\pm$ $<$0.01 & 0.38 $\pm$ $<$0.01 & 1.37 $\pm$ 0.11 & 3.03 $\pm$ 0.81 & 4.63 $\pm$ 1.74 & 16.11 $\pm$ 7.39 \\
        SDSSJ0839+0454 & 0.18 $\pm$ $<$0.01 & 0.26 $\pm$ 0.01 & 1.00 $\pm$ 0.12 & 1.14 $\pm$ 0.98 & 7.13 $\pm$ 2.84 & 27.83 $\pm$ 4.99 \\
        SDSSJ0943-0100 & 0.63 $\pm$ $<$0.01 & 1.05 $\pm$ $<$0.01 & 4.62 $\pm$ 0.10 & 9.11 $\pm$ 0.74 & 26.76 $\pm$ 2.83 & 62.98 $\pm$ 5.46 \\
        
		\hline
	\end{tabular}
\end{table*}

The combined photometry (WISE + PACS + SPIRE) is modelled with a spectral energy distribution (SED) comprising two components - a quasar template, accounting for the hot dust arising from quasar heating, and a star-forming model tracing the cooler dust emission. The selected quasar template is taken from the work of \cite{mor12}, as provided by \cite{lani17} and is based on the intrinsic SEDs of 115 nearby Type-1 AGN, spanning luminosities L$_{5100}$ $\simeq$ 10$^{43.2}$-10$^{45.9}$ ergs$^{-1}$. The star-forming component of the model is characterised by a modified black body (or \textit{greybody}) curve, $S(\nu)$, \citep{casey12} of the form;

\begin{equation}
S(\nu) = (1 - e^{\left[\frac{-\nu}{\nu_{0}} \right] ^{\beta}}) \times \frac{\nu^3}{e^{\frac{h\nu}{kT}}-1},
\label{eqn:Sv_full}
\end{equation}

where $\nu_{0}$ is the frequency ($\nu$) at which optical depth is equal to unity \citep{draine06}. T$_{\rm{DUST}}$ and $\beta$ are the dust temperature and the emissivity index respectively. The combined photometry is simultaneously fit with a combination of the two components (hot torus + star-forming galaxy) to create a total model describing the observed NIR to FIR photometry of the LoBAL sample i.e. 

\begin{equation}
M_{\rm{TOT}} = (X_{\rm{TORUS}} \times M_{\rm{TORUS}}) + (X_{\rm{SF}} \times S(\nu)),
\label{eqn:ModTot}
\end{equation}

where the total model, M$_{\rm{TOT}}$, is given to be the sum of the torus, M$_{\rm{TORUS}}$, and the star-forming, S($\nu$), models, scaled by the factors X$_{\rm{TORUS}}$ and X$_{\rm{SF}}$ respectively. 

\begin{figure*}
    \centering 
	\includegraphics[trim= 50 10 50 20 ,clip,width=\textwidth]{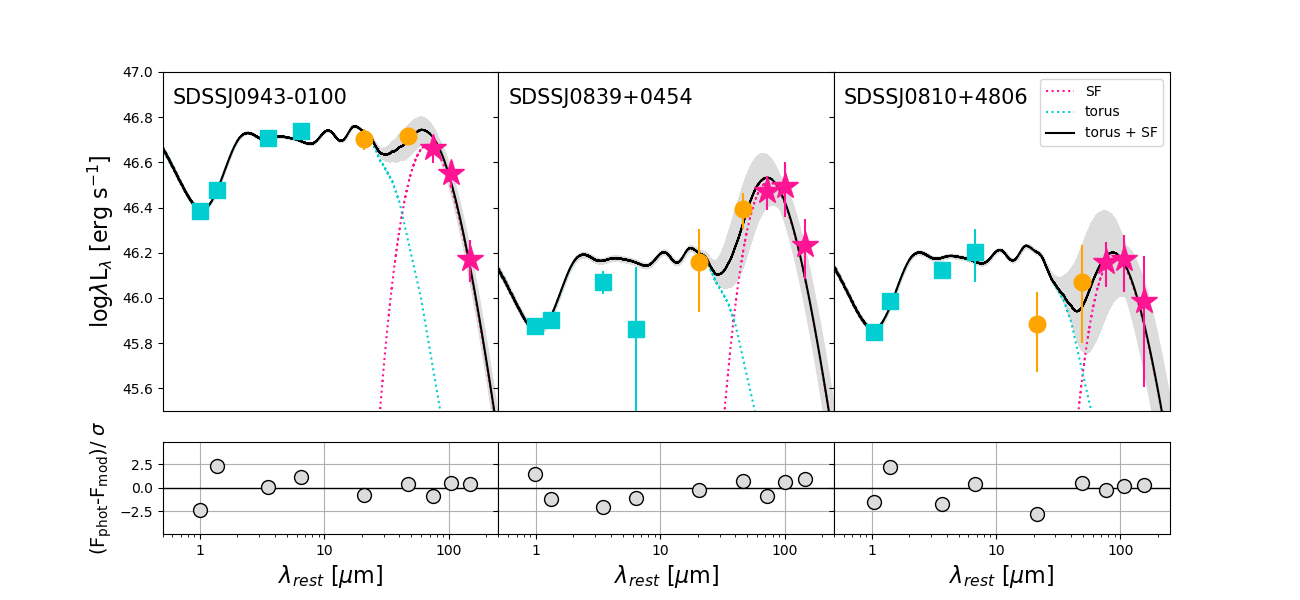}
\caption{\textbf{\textit{Upper:}} Best-fit SED template based on the combined WISE (\textit{blue squares}) + PACS (\textit{orange circles}) + SPIRE (\textit{pink stars}) photometry. The total model (\textit{black solid line}) is comprised of contributions from a hot torus (\textit{cyan dotted line}) and a star forming galaxy (\textit{pink dotted line}). Grey shaded regions denotes the 1$\sigma$ uncertainty in the total model, based on the derived uncertainties on the scaling factors (X$_{\rm{TORUS}}$, X$_{\rm{SF}}$) and dust temperature (T$_{\rm{DUST}}$). \textbf{\textit{Lower:}} Error weighted residuals of the best-fit model.}
\label{fig:SEDs}
\end{figure*}

\begin{figure*}
	\centering 
	
	\subfigure{\includegraphics[trim= 10 0 28 28 ,clip,width=.32\textwidth]{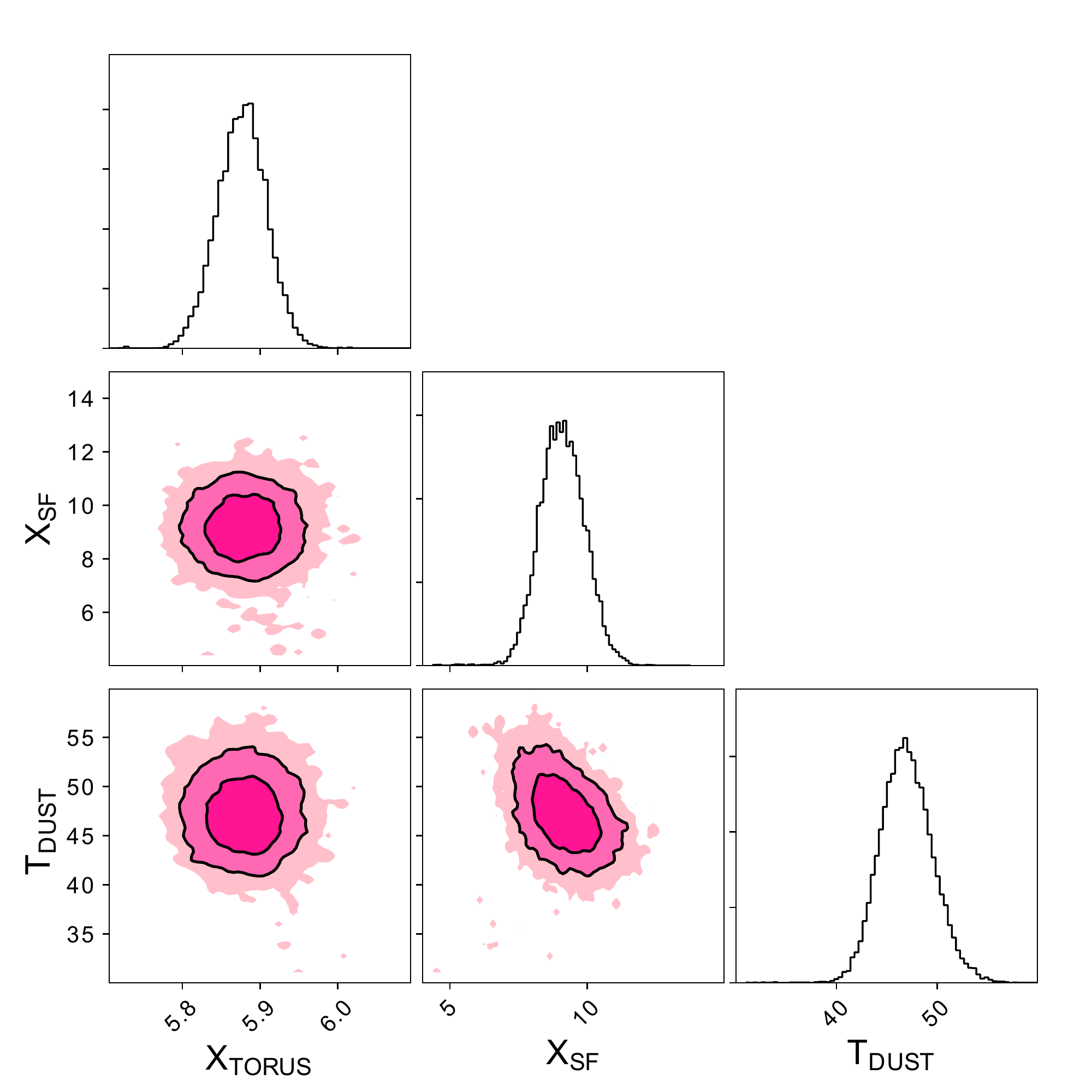}}
	\subfigure{\includegraphics[trim= 10 0 28 28 ,clip,width=.32\textwidth]{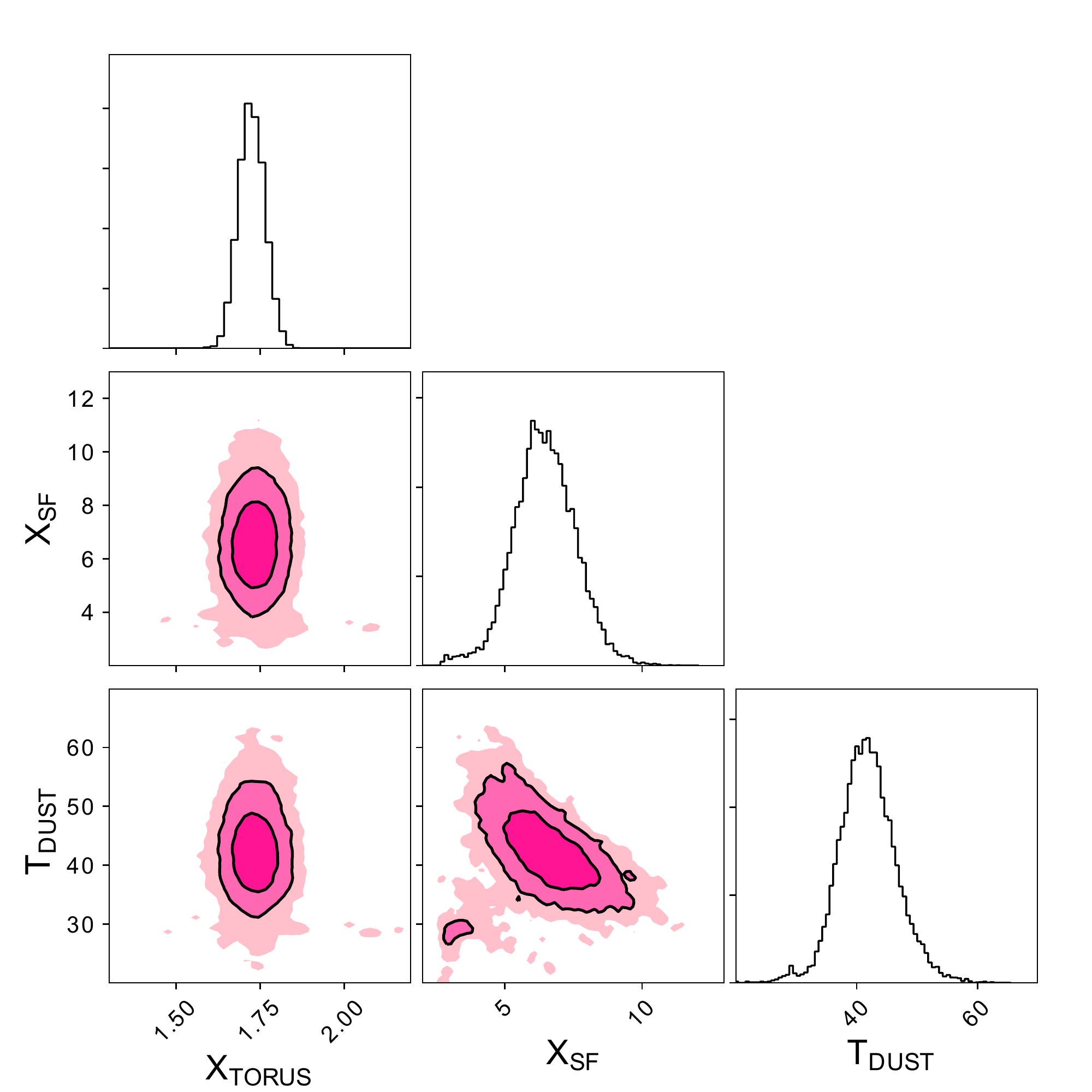}}
	\subfigure{\includegraphics[trim= 10 0 28 28 ,clip,width=.32\textwidth]{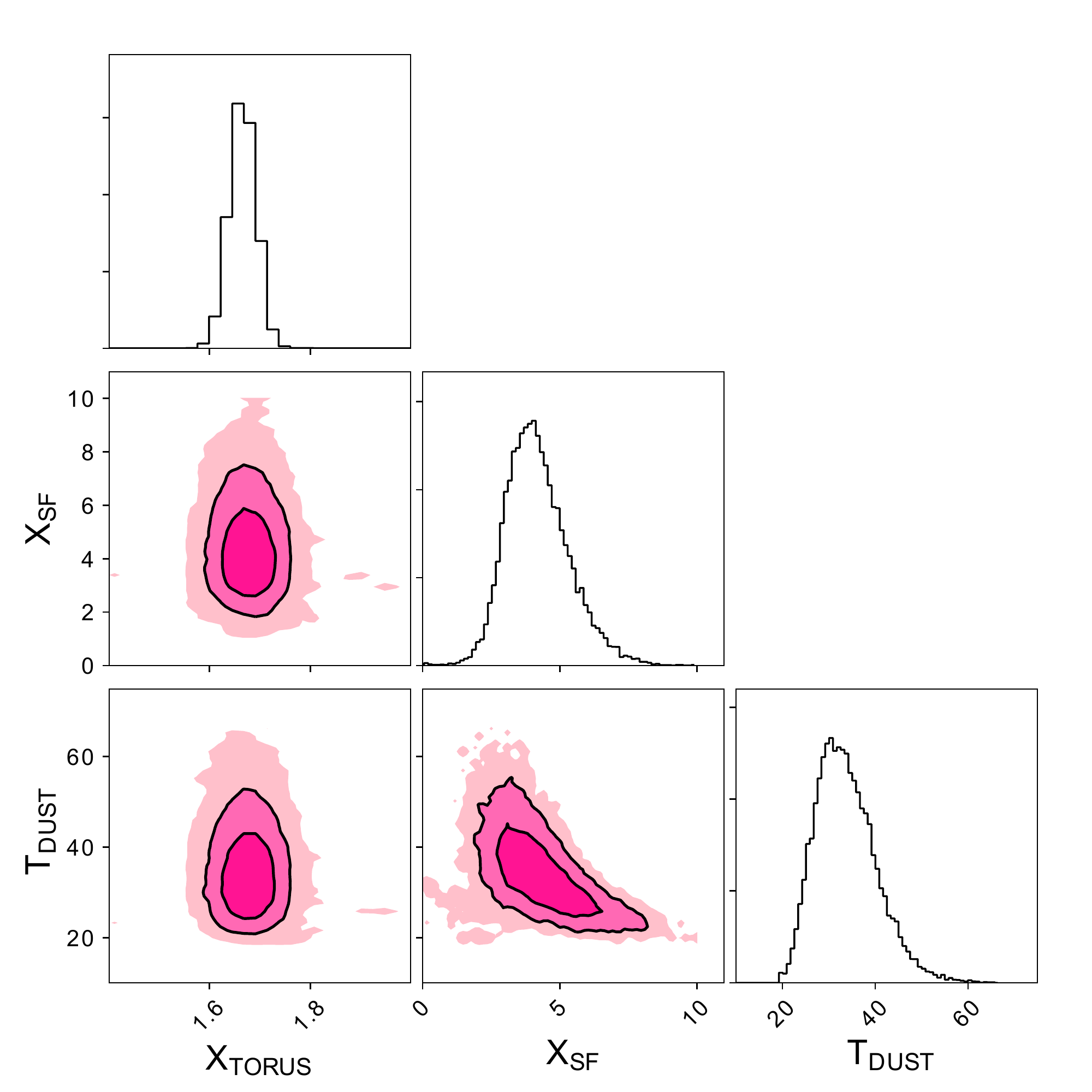}}
	
\caption{1D and 2D parameter solutions for the SED fitting in Fig.~\ref{fig:SEDs}. Contours denote the 1, 2 and 3$\sigma$ confidence bounds on the derived parameter values. }

\label{fig:CornerPlot}
\end{figure*}

The fitting routine utilises a Markov-Chain Monte-Carlo (MCMC) method in order to obtain full posterior distributions on the best-fit model parameters and to marginalise over any nuisance parameters. Throughout the fitting the vertical scaling of both the torus SED (X$_{\rm{TORUS}}$) and the greybody template (X$_{\rm{SF}}$), and the dust temperature of the greybody template (T$_{\rm{DUST}}$) are set as free parameters. Given the limited photometry tracing the cool dust emission, we adopt a fixed value of $\beta$ = 1.6, consistent with the work of \cite{priddey01}, although we note that initial tests in which $\beta$ was included as an additional free parameter demonstrated it to have a negligible impact on the inferred SFRs of our LoBAL sample. Boundaries on each of the free parameters are set such that the dust temperature, T$_{\rm{DUST}}$, is constrained within 20-70K, spanning the full range of observed dust temperatures for quasar hosts at high redshift \citep[e.g.][]{casey12}. Similarly, boundaries on the vertical scaling factors, X$_{\rm{TORUS}}$ and X$_{\rm{SF}}$, are set to range 0.1-10, following an initial normalisation. Flat priors are assumed throughout. The results of the fitting are given in Fig.~\ref{fig:SEDs} along with the best-fit model residuals, whilst the corresponding 1D and 2D parameter solutions are presented in Fig.~\ref{fig:CornerPlot}. 

Based on Fig.~\ref{fig:SEDs}, the quasar contamination at $\lambda$ $>$250$\mu$m appears negligible in the detected LoBAL targets, comprising $<$10 per cent of the total flux in the \textit{Herschel} SPIRE bands. Fig.~\ref{fig:CornerPlot} also indicates a possible degeneracy between the the scaling of the greybody template, X$_{\rm{SF}}$, and the inferred dust temperature, T$_{\rm{DUST}}$, with lower scaling factors yielding lower temperatures. In general, however, the fitting routine returns well constrained values for both the dust temperature and the scaling of the two model components. Based on the best-fit parameters, we integrate over the star-forming component of the model (Fig.~\ref{fig:SEDs}) from 8-1000$\mu$m \citep{kennicutt12} to estimate the FIR luminosity of the LoBAL host galaxy, L$_{\rm{FIR}}$;

\begin{equation}
L_{\rm{FIR}} = 4\pi D_{\rm{L}}^{2} \int_{\nu_{\rm{min}}}^{\nu_{\rm{max}}} S(\nu)  d\nu, 
\label{eqn:l_fir}
\end{equation}

where $D_{\rm{L}}$ is the luminosity distance (in cm) and $\nu_{\rm{min}}$, $\nu_{\rm{max}}$ denote the FIR integral limits in terms of frequency ($\nu_{\rm{min}}$ = 0.3 THz; $\nu_{\rm{max}}$ = 37.5 THz). SFRs are then estimated from L$_{\rm{FIR}}$ using the relation outlined in \cite{kennicutt12}, which states;

\begin{equation}
SFR = 4.5\times10^{-44} \times L_{\rm{FIR}},
\label{eqn:lSFR}
\end{equation}

where the resulting SFR is given in units of M$_{\rm{\odot}}$yr$^{-1}$. Based on this conversion (Eqn.~\ref{eqn:lSFR}), we derive SFRs of 740$_{-170}^{+220}$, 1610$_{-260}^{+280}$ and 2380$_{-210}^{+220}$ M$_{\odot}$yr$^{-1}$ for SDSSJ0810+4806, SDSSJ0839+0454 and SDSSJ0943-0100 respectively (Table~\ref{tab:SFRs}), finding evidence for prolific star formation in our detected LoBAL sub-sample. This result is consistent with the work of \cite{pitchford19}, who also find evidence for prolific star formation ($\sim$2000M$_{\odot}$yr$^{-1}$) in an iron LoBAL (FeLoBAL) at z=1.046, associated with recent starburst activity in the galaxy.

\begin{table}
	\centering
	\caption{The inferred dust temperatures, T$_{\rm{DUST}}$, the FIR luminosity (8-1000$\mu$m), L$_{\rm{FIR}}$ and SFR estimates for the sub-sample of detected LoBALs and for the stacked non-detections, where the 3$\sigma$ upper limit on L$_{\rm{FIR}}$ and the SFR is instead given. Quoted uncertainties denote the 1$\sigma$ error derived within the fitting routine.}
	\label{tab:SFRs}
	\begin{tabular}{lccc} 
	        \hline
        Name & \multicolumn{1}{p{1.5cm}}{\centering T$_{\rm{DUST}}$ \\ $[$K$]$} & \multicolumn{1}{p{1.4cm}}{\centering log L$_{\rm{FIR}}$ \\ $[$ergs$^{-1}$ $]$} & \multicolumn{1}{p{1.4cm}}{\centering SFR$_{\rm{FIR}}$ \\ $[$M$_{\odot}$yr$^{-1}$ $]$} \\
        \hline
        
    SDSSJ0810+4806  & 33.49$_{-5.70}^{+7.11}$  & 46.21$_{-0.12}^{+0.11}$ & 740$_{-170}^{+220}$  \\
    SDSSJ0839+0454  & 42.02$_{-4.32}^{+4.64}$  & 46.55$_{-0.08}^{+0.07}$ & 1610$_{-260}^{+280}$  \\
    SDSSJ0943 -0100 & 47.07$_{-2.50}^{+2.67}$  & 46.72$_{-0.04}^{+0.04}$ & 2380$_{-210}^{+220}$  \\
    Stacked non-detections & NaN & $<$46.04 & $<$440\\
		\hline
	\end{tabular}
\end{table}

To explore the average star-forming properties of our LoBAL sample, we estimate the SFR of the stacked non-detections (Fig.~\ref{fig:ImgStack}), based on the aperture fluxes derived in Sec.~\ref{sec:ImgStack}. We note that whilst it is not possible to constrain the potential quasar contamination in the same manner as for the detected targets, we have demonstrated that the contamination caused by quasar heating to the \textit{Herschel} SPIRE bands is likely to be minimal (Fig.~\ref{fig:SEDs}). We therefore fit the stacked aperture photometry with a single greybody component, setting the dust temperature, T$_{\rm{DUST}}$, and the overall scaling of the model, X$_{\rm{SF}}$ as free parameters. Based on the best-fit model, we again integrate over the greybody template from 8-1000$\mu$m to calculate the total FIR luminosity (Eqn.~\ref{eqn:l_fir}) and thus estimate the upper limit on the SFR (Eqn.~\ref{eqn:lSFR}) of the non-detections. As such, we derive a 3$\sigma$ upper limit of $<$440 M$_{\odot}$yr$^{-1}$ on the SFR of the non-detections, meaning even though these targets remain undetected with \textit{Herschel} SPIRE, we cannot rule out moderate to high SFRs in these systems, albeit lower than the rates derived for the detected LoBAL targets.


\section{Discussion}
\label{sec:Discussion}

\subsection{Comparison of FIR Properties}
\label{sec:CompSamples}

Having found evidence for prolific star formation in individual LoBALs in our sample, we now seek to compare the FIR emission of LoBALs to that of both HiBAL and non-BAL quasar populations. If LoBALs mark a post-starburst phase in the lifetime of a quasar, as predicted by the evolutionary paradigm \citep{boroson92}, we would expect the FIR luminosity of our sample to exceed that of HiBALs and non-BALs, neither of which are typically associated with starburst activity in the host. 

\subsubsection{LoBALs vs. HiBALs}
\label{sec:CompHB}

\begin{figure}
	\centering 
	\includegraphics[trim= 0 10 30 20 ,clip,width=.5\textwidth]{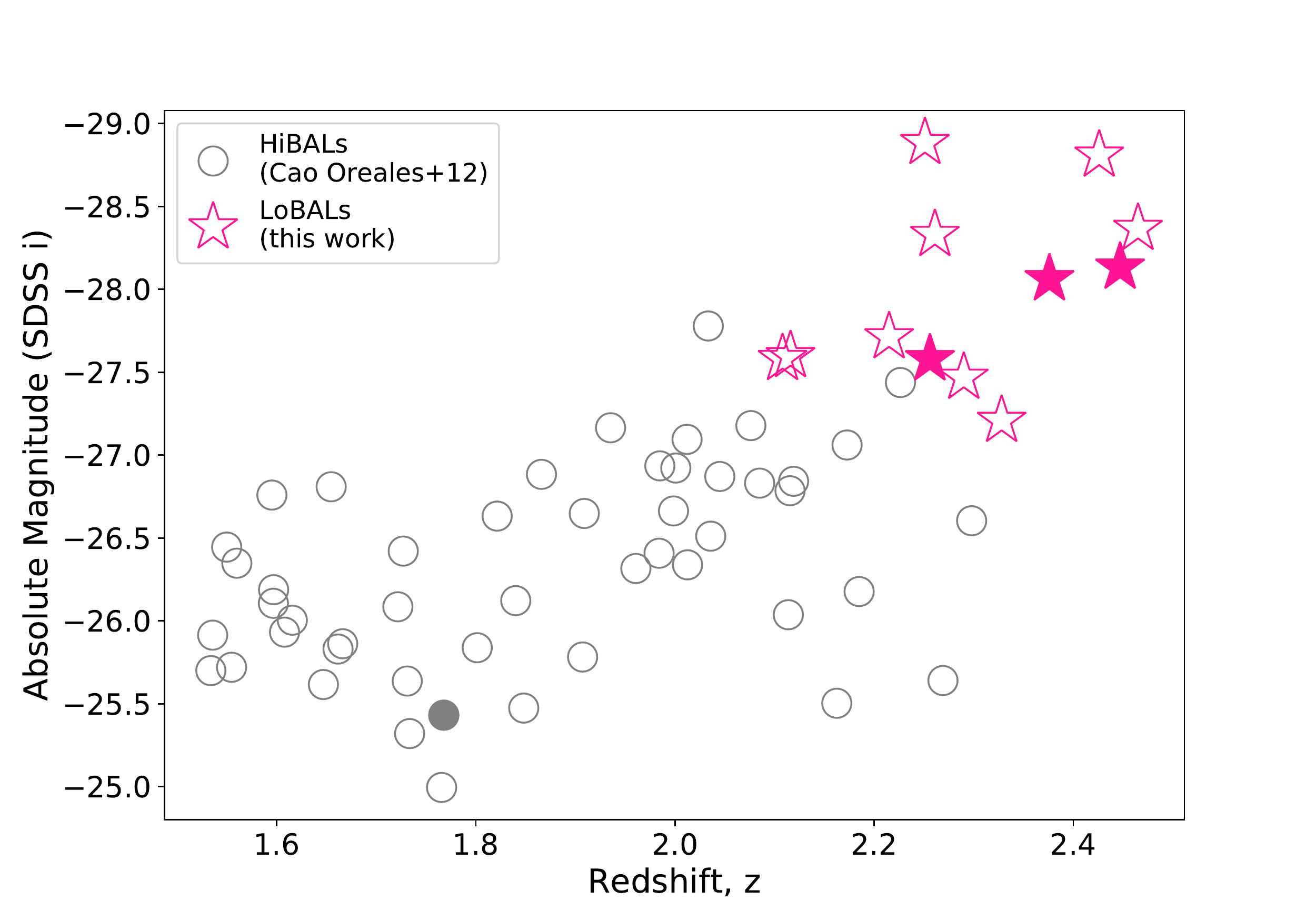}
\caption{A comparison of the HiBAL sample in \protect\cite{cao12} (\textit{grey circles}) and the LoBAL sample of this paper (\textit{pink stars}) in terms of their SDSS $i$-band magnitude and redshift. Filled symbols denote targets with aperture fluxes above $>$33.5, $>$37.7 and $>$44.0 mJy beam$^{-1}$ at 250, 350 and 500$\mu$m respectively, corresponding to the 5$\sigma$ detection thresholds in \protect\cite{cao12}.}
\label{fig:HiBALcompare}
\end{figure}

Numerous studies have found HiBALs to be consistent with an orientation model of BALs, in which all quasars have BAL winds, but can only be observed as such along particular sight lines \citep[e.g.][]{weymann91,reichard03,willott03,gallagher07,cao12}. Unlike HiBALs however, LoBALs may instead mark a short-lived evolutionary phase in the lifetime of quasars. A key prediction of this evolutionary paradigm is the enhancement of star formation among LoBALs, indicating recent or ongoing starburst activity. Already in Sec.~\ref{sec:SFRs}, we have demonstrated the three detected LoBALs at 2.0$<$ z $<$2.5 to exhibit high SFRs, but now we seek to compare these to those of HiBALs, by looking for differences in the FIR detection rates of the two populations. To this end, we consider a sample of 49 HiBALs presented in \cite{cao12}. The sample was initially selected from SPIRE imaging data at 250, 350 and 500$\mu$m as part of the \textit{Herschel} Astrophysical Teraherz Large-Area Survey (H-ATLAS) and is presented alongside our LoBAL targets in Fig.~\ref{fig:HiBALcompare}. 

\begin{figure}
	\centering 
	\includegraphics[trim= 0 35 10 20 ,clip,width=.5\textwidth]{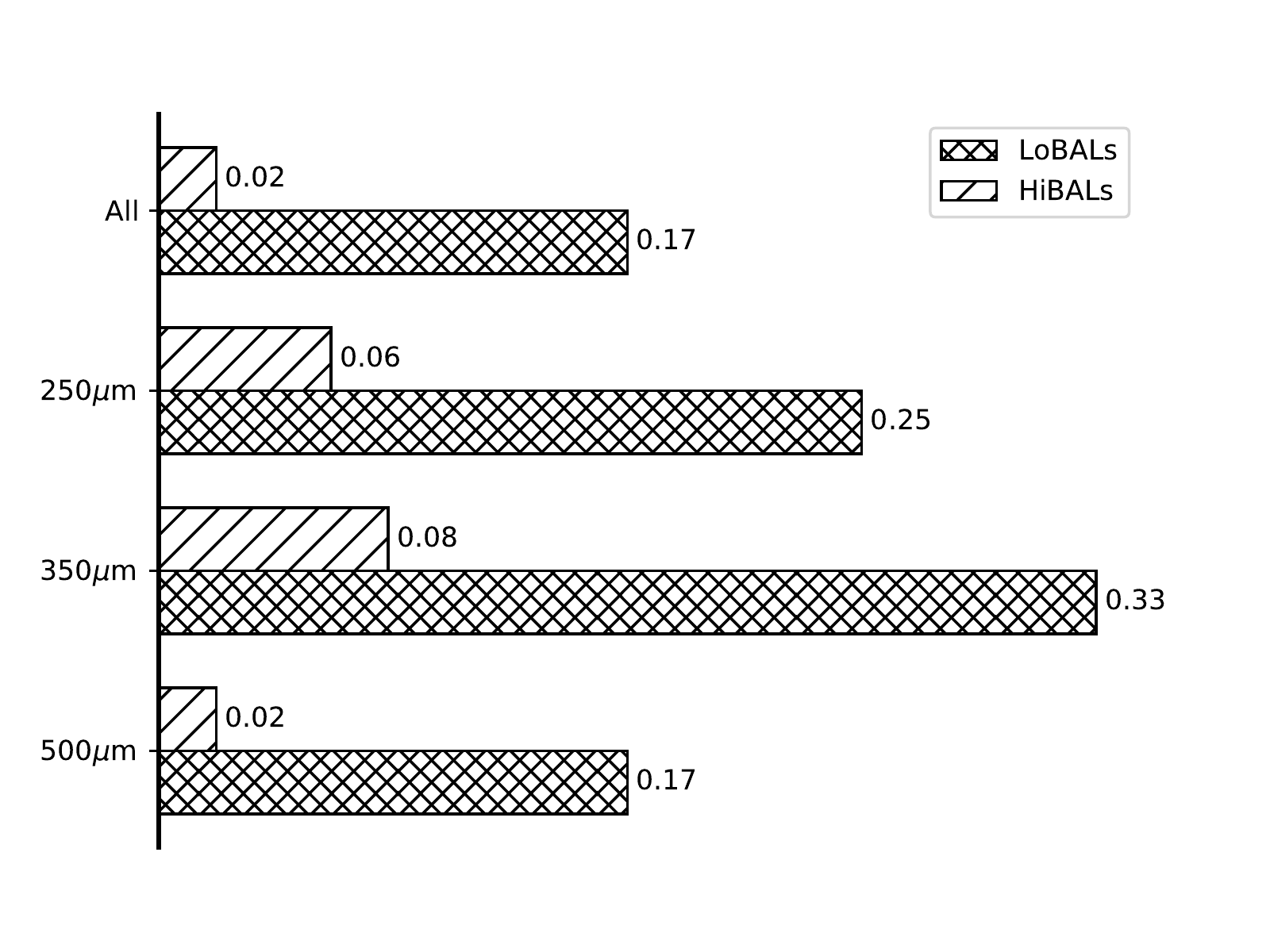}
\caption{Fraction of the LoBAL sample in each of the SPIRE bands with aperture fluxes above the 5$\sigma$ detection thresholds of HiBALs in \protect\cite{cao12}($>$33.5, $>$37.7 and $>$44.0 mJy beam$^{-1}$ at 250, 350 and 500$\mu$m respectively).}
\label{fig:loHi}
\end{figure}

Of the 49 HiBALs in \cite{cao12}, one is detected above their 5$\sigma$ flux threshold in all three SPIRE bands (2 per cent), corresponding to $>$33.5, $>$37.7 and $>$44.0 mJy beam$^{-1}$ at 250, 350 and 500$\mu$m respectively. To compare this detection rate with that of our LoBAL sample, we count the number of LoBALs with aperture fluxes (Tab.~\ref{tab:DataTab}) lying above the flux thresholds of \cite{cao12} in each band. Three targets in our sample have S$_{250}$ $>$ 33.5mJy, four have S$_{350}$ $>$ 37.7mJy and two have S$_{500}$ $>$ 44.0mJy. Only two of the 12 LoBALs in our sample (SDSSJ0943-0100 and SDSSJ0839+0454) lie above the 5$\sigma$ flux thresholds of \cite{cao12}  in every band, corresponding to 17 per cent of the sample. We highlight that this is lower than the fraction of LoBALs detected at $>$5$\sigma$ (25 per cent) in Sec.~\ref{sec:AperPhot}, due to the different detection criteria of this work and that of \cite{cao12}. We therefore find an enhancement in the fraction of LoBALs with fluxes above the detection 5$\sigma$ thresholds of \cite{cao12} by a factor of $\sim$8.5. Later in Sec.~\ref{sec:caveats} we discuss the potential implications of the observed redshift and luminosity variation between the samples on this inferred enhancement, but for now we simply note an increase in the FIR fluxes of LoBALs compared to HiBALs across all \textit{Herschel} SPIRE bands (Fig.~\ref{fig:loHi}). Using a binomial probability distribution, we test the significance of this apparent enhancement given the relatively small sample sizes by calculating the $p$-value statistic. When considering targets with aperture fluxes above the detection thresholds of \cite{cao12} in all bands, we derive a $p$-value of 0.09, ruling out the two samples being drawn from the same underlying population with a confidence of $>$90 per cent. Although this apparent enhancement is observed across all the \textit{Herschel} SPIRE bands, it appears particularly evident at 350$\mu$m, where we find a third of our LoBAL sample (33 per cent) with S$_{350}$ $>$ 37.7mJy compared to just 6 percent of HiBAL targets and thus conclude an enhancement in the FIR emission of LoBALs with a confidence $>$99.9 per cent in this band. Our results therefore support an enhancement in the FIR luminosity of LoBALs with regard to the HiBAL population, potentially indicative of higher dust masses among LoBALs, associated with active star formation in the galaxy.

\subsubsection{LoBALs vs. non-BALs}
\label{sec:CompNB}

\begin{figure}
	\centering 
	\includegraphics[trim= 0 10 30 20 ,clip,width=.5\textwidth]{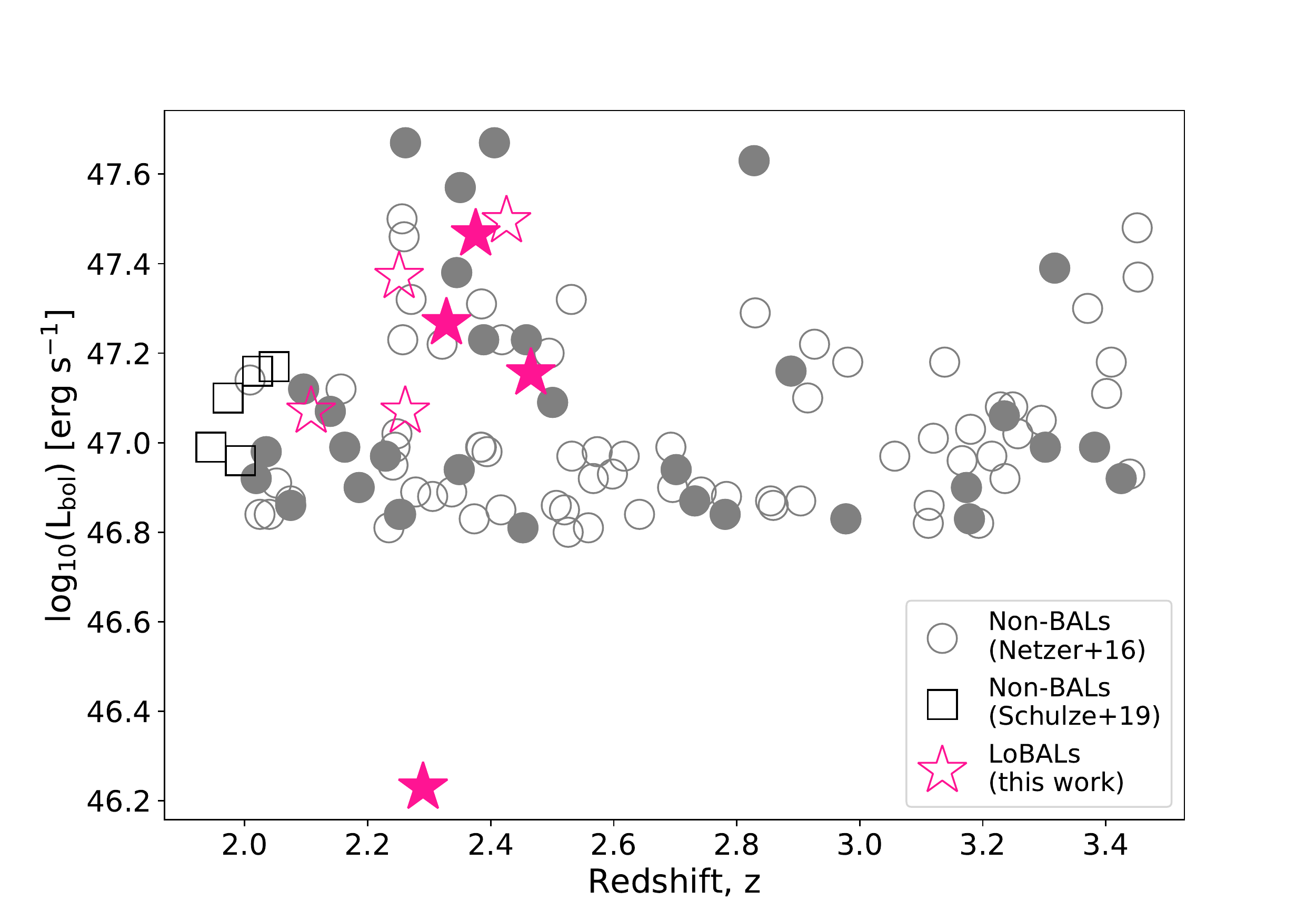}
\caption{Plot showing the non-BAL samples from \protect\cite{netzer16} (\textit{circles}) and \protect\cite{schulze19} (\textit{squares}) in terms of their bolometric quasar luminosity (L$_{\rm{bol}}$) and redshift. The eight targets in our LoBAL sample with spectral information on L$_{\rm{bol}}$ are plotted as reference (\textit{pink stars}). Filled symbols denote targets with aperture fluxes in all three SPIRE bands above the 3$\sigma$ thresholds of \protect\cite{netzer16} ($>$17.4, $>$18.9 and $>$20.4 at 250, 350 and 500$\mu$m respectively).}
\label{fig:nonBALcompare}
\end{figure}

\begin{figure}
	\centering 
	\includegraphics[trim= 0 35 10 20 ,clip,width=.5\textwidth]{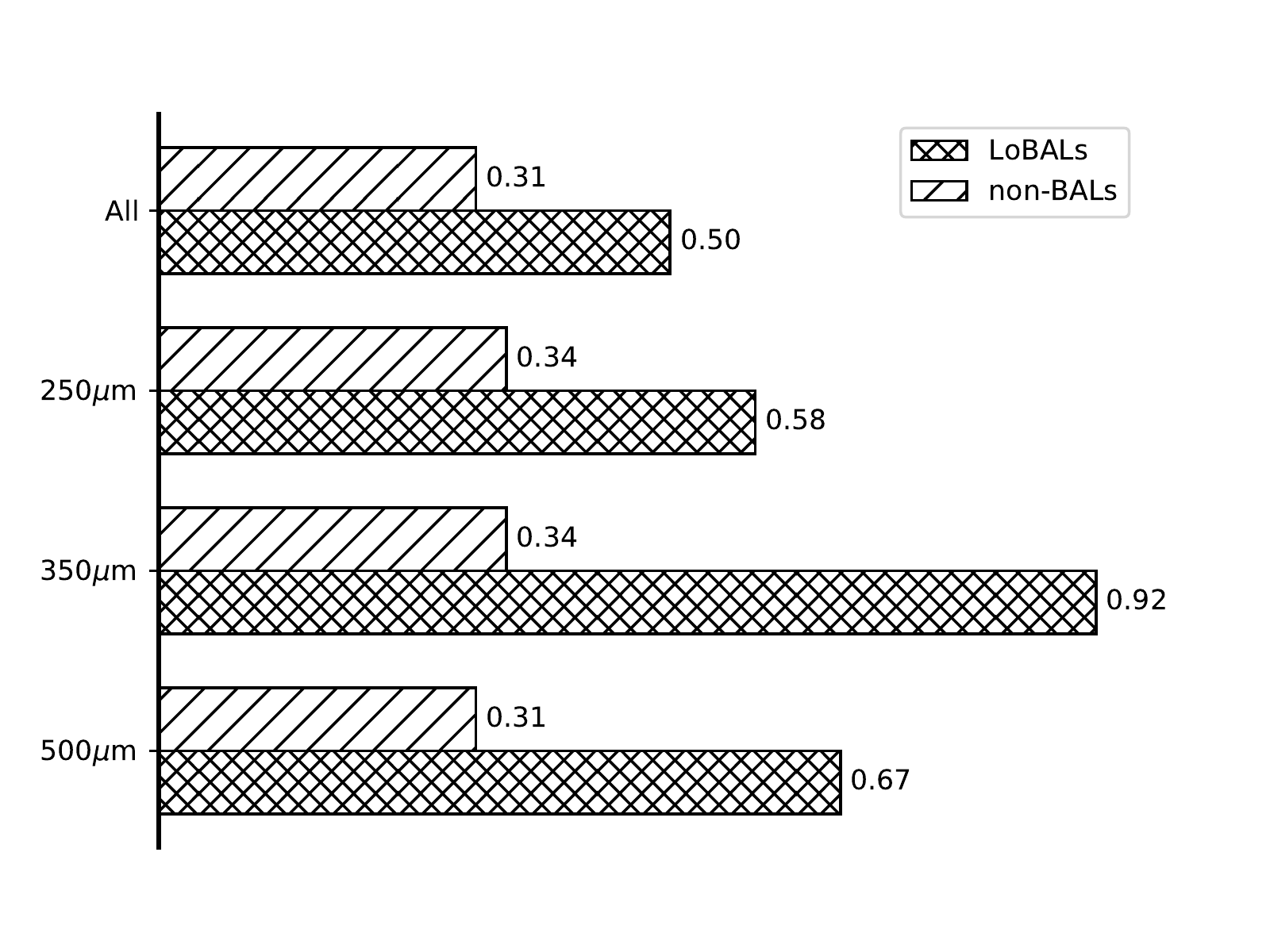}
\caption{Fraction of the LoBAL sample in each of the SPIRE bands with aperture fluxes above the 3$\sigma$ detection thresholds of non-BALs in \protect\cite{netzer16}($>$17.4, $>$18.9 and $>$20.4 at 250, 350 and 500$\mu$m respectively).}
\label{fig:loNon}
\end{figure}

A recent study by \cite{schulze17} concludes that z$\sim$2 LoBALs are entirely consistent with non-BALs in terms of their black hole mass, M$_{\rm{BH}}$, Eddington luminosity, L$_{\rm{Edd}}$, and rest-frame optical spectra, finding no evidence that these galaxies represent a special phase in quasar evolution. Their study however, draws no conclusions on whether LoBALs exhibit an enhancement in star formation - a key prediction of the evolutionary paradigm which we seek to test here. To this end, we compare our LoBAL sample to the \textit{Herschel} SPIRE imaging of 100 luminous (L$_{\rm{bol}}$ $>$ 46.5 erg s$^{-1}$) type-1 non-BAL quasars from the work of \cite{netzer16}. All non-BAL targets in \cite{netzer16} were optically selected from data release 7 (DR7) of the Sloan Digital Sky Survey (SDSS) and spectroscopically confirmed to lie at 2.0$<$ z $<$3.5. This non-BAL sample is presented in Fig.~\ref{fig:nonBALcompare} alongside the eight targets in our LoBAL sample for which luminosity information is available, although we highlight that comparisons are made based on our full LoBAL sample (12 targets). Whilst we cannot directly compare L$_{\rm{bol}}$ of the remaining four targets in our LoBAL sample, we note that the redshifts of our entire sample lie within the range of redshifts in \cite{netzer16}. Of the 100 non-BAL quasars \citep{netzer16}, 31 (31 per cent) are detected above a 3$\sigma$ threshold in all three of the SPIRE bands, nominally corresponding to a flux threshold of $>$17.4, $>$18.9 and $>$20.4mJy at 250, 350 and 500$\mu$m respectively. Comparatively, six of the 12 LoBALs in our sample (50 per cent) are found to lie above these same flux limits in all SPIRE bands, indicating an enhancement of factor $\sim$1.5 with regards to the non-BAL population (Fig.~\ref{fig:loNon}). We test the significance of this apparent enhancement by calculating the $p$-value statistic, returning $p$ = 0.09 and thus finding the FIR properties of the two populations to be independent with a confidence of $>$90 per cent. 

Furthermore, when we compare the FIR properties of our LoBAL targets to a more recent sample of luminous non-BAL quasars at z$\sim$2 \citep{schulze19} we find a similar enhancement in the FIR fluxes of LoBALs. The work of \cite{schulze19} outlines observations from the Atacama Large Millimeter Array (ALMA) for 20 non-BALs, of which five have also been imaged with \textit{Herschel} SPIRE. Of these five targets, none are detected above their 5$\sigma$ flux threshold (250$\mu$m;$>$33.5mJy, 350$\mu$m;$>$37.7mJy, 500$\mu$m;$>$44.0mJy) in every band, returning a $p$-value of $p$ = 0.06 when compared to the number of LoBALs in our sample with aperture fluxes above the same thresholds. We highlight that this is likely an overestimation of the statistical difference between the two samples given the small sample size and lack of non-BAL detections, but nevertheless find our results to indicate an enhancement in the FIR fluxes of LoBALs, consistent with the comparisons made to the \cite{netzer16} sample. 

\cite{schulze19} additionally derive SFRs for their full sample of 20 non-BALs based on their 850$\mu$m fluxes in ALMA. This allows us to directly compare these rates to the SFRs inferred at the sensitivity limit of our sample and therefore determine whether the observed enhancement in the FIR detection rate is indeed associated with higher SFRs in LoBALs. As such, we scale a greybody template to a single photometry point denoting the average 5$\sigma$ flux threshold of our sample at 250$\mu$m, corresponding to 25.39mJy. The scaling utilises a basic $\chi^2$ minimisation, assuming fixed parameters matching those in \cite{schulze19} (i.e. T$_{\rm{DUST}}$=47K and $\beta$=1.6) and varying only the normalisation of the curve. L$_{\rm{FIR}}$ is derived from the fitted template by integrating over the FIR wavelengths, following the methods outlined in Sec.~\ref{sec:SFRs}, and converted to a SFR using Eqn.~\ref{eqn:lSFR}. This returns SFR = 640M$_{\odot}$yr$^{-1}$ at the flux sensitivity of our sample. Just one of the 20 non-BALs in \cite{schulze19} exhibit SFR $\geq$ 640M$_{\odot}$yr$^{-1}$ (5 per cent), compared to three LoBALs in our sample (25 per cent). Based on this result, we conclude that we not only see evidence for an enhancement in the FIR fluxes of our sample, but also that we find direct evidence for the enhancement of star formation among LoBALs.

\subsection{LoBAL Environments}
\label{sec:LoBALEnv}

\begin{figure}
	\centering 
	\subfigure{\includegraphics[trim= 80 23 40 25,clip,width=.5\textwidth]{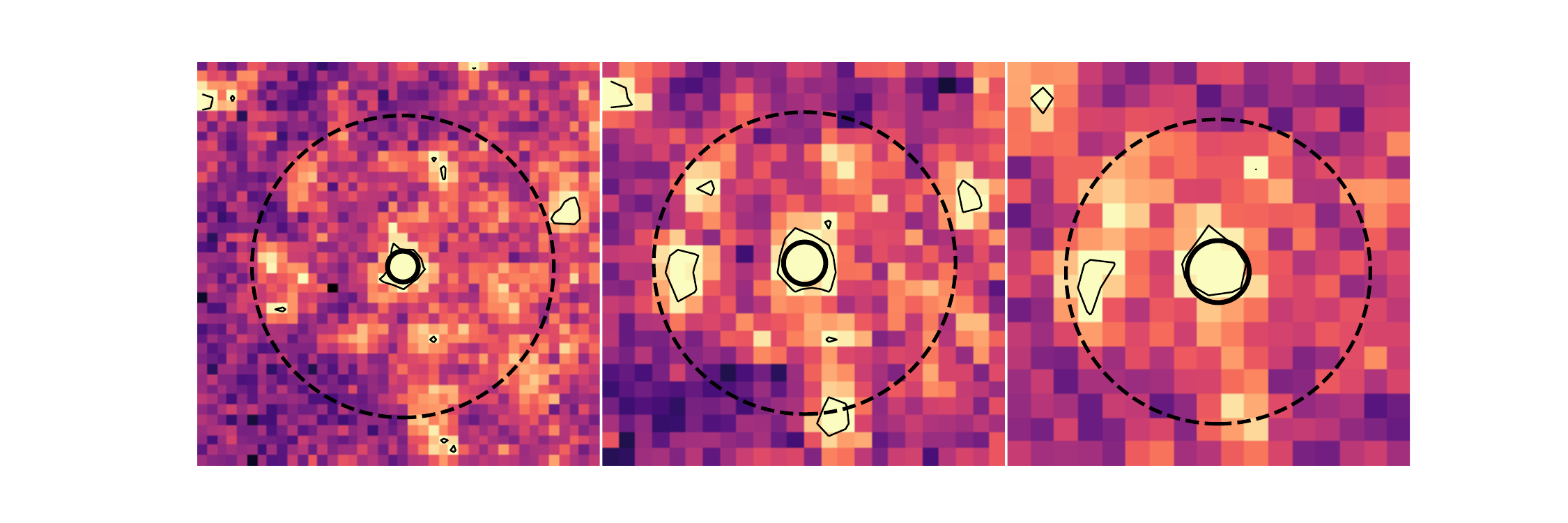}}
	
	\subfigure{\includegraphics[trim= 45 100 10 0,clip,width=.5\textwidth]{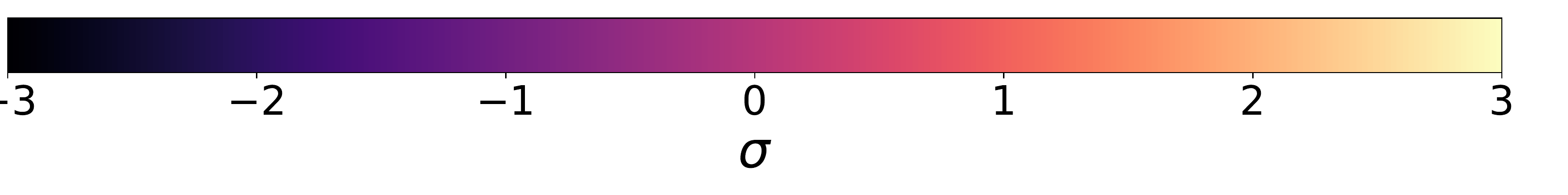}}
	
\caption{S/N maps and 5$\sigma$ contours mapping serendipitous sources in the SPIRE images of SDSS0943-0100. Dotted line denotes the 1.5 arcmin radius area over which sources were identified.}

\label{fig:env_example}
\end{figure}

Having found evidence for enhanced SFRs in our LoBAL sample, we now investigate the environments in which these systems reside. If LoBALs comprise an evolutionary quasar phase following a merger-induced starburst, we may expect LoBALs to exist in denser environments associated with more frequent galaxy-galaxy interactions. To this end, we look for any serendipitous detections in the SPIRE images within a 1.5 arcmin radius of our targets ($\sim$1Mpc scales), making use of the S/N maps derived in Sec.~\ref{sec:AperPhot} (see Fig.~\ref{fig:env_example}). We select the same 5$\sigma$ detection threshold as our target detections in order to exclude spurious sources in the image. To minimise any potential contamination from bad image pixels, sources are further required to span multiple pixels in the image. Aperture fluxes for each serendipitous source are derived following the methods outlined in Sec.~\ref{sec:AperPhot}, centering the appropriate aperture on the brightest pixel in the source. The number of sources detected in each band for each target (including LoBAL target detections) are given in Table~\ref{tab:nCounts}. We note that the number of detected sources at 500$\mu$m is a factor $\sim$4 lower than at 250 and 350$\mu$m, but suggest this may be due to the large pixel scale in this band (14 arcsec/pixel) and the criteria by which we select sources in the image.

\begin{table}
	\centering
	\caption{Number counts for the serendipitous detections within 1.5 arcmin of each LoBAL target.}
	\label{tab:nCounts}
	\begin{tabular}{lccc} 
		\hline
		Name & N$_{250}$ ($>$5$\sigma$) & N$_{350}$ ($>$5$\sigma$) & N$_{500}$ ($>$5$\sigma$) \\
		\hline
        SDSSJ0753+2102 &  0 & 2 & 0 \\
        SDSSJ0810+4806 &  3 & 4 & 1 \\
        SDSSJ0839+0454 &  4 & 3 & 1 \\
        SDSSJ0943-0100 &  5 & 5 & 2 \\
        SDSSJ0957+5120 &  0 & 0 & 0 \\
        SDSSJ1011+5155 &  0 & 0 & 0 \\
        SDSSJ1028+5110 &  3 & 2 & 0 \\
        SDSSJ1132+0104 &  1 & 4 & 0 \\
        SDSSJ1341-0036 &  2 & 0 & 0 \\
        SDSSJ1352+4239 &  2 & 2 & 1 \\
        SDSSJ1516+0029 &  2 & 5 & 1 \\
        SDSSJ1723+5553 &  0 & 0 & 0 \\
        \hline
        \textbf{Total:} & \textbf{22} & \textbf{27} & \textbf{6} \\
	\end{tabular}
\end{table}

\begin{figure*}
	\centering 
	
	\includegraphics[trim= 64 0 95 20 ,clip,width=\textwidth]{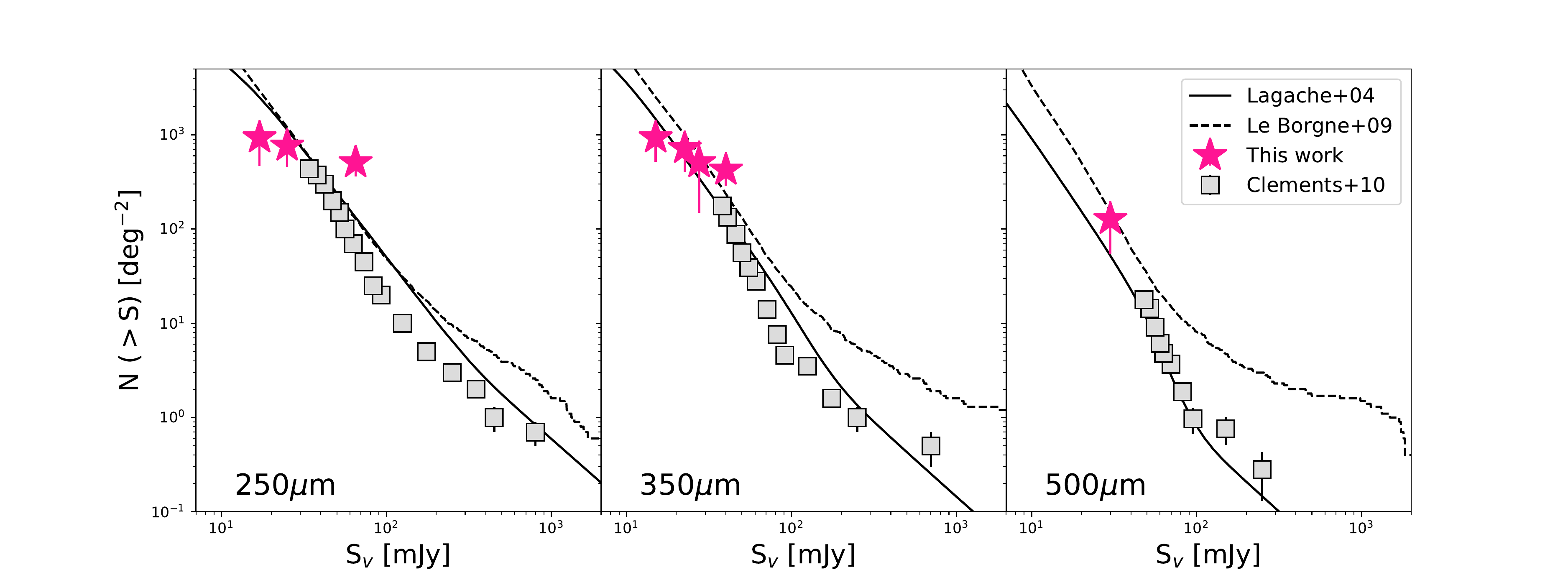}
	
\caption{Number counts for the $>$5$\sigma$ detections within 1.5 arcmin of our LoBAL targets compared to those of the H-ATLAS field given in \protect\cite{clements10} (\textit{grey squares}). Model predictions from \protect\cite{lagache04} (\textit{solid line}) and \protect\cite{le09} (\textit{dashed line}) are included for reference.}
\label{fig:nCounts}
\end{figure*}

To test whether the detected sources in Table~\ref{tab:nCounts} indicate an overdensity in the environments of our LoBAL sample, we directly compare the LoBAL number counts to blank field counts presented in \cite{clements10}, which are based on the first 14 deg$^2$ of the Herschel-ATLAS survey catalogue. Fig.~\ref{fig:nCounts} plots the cumulative frequency of the LoBAL source counts as a function of their flux at 250, 350 and 500$\mu$m, finding them to be entirely consistent with the blank field number counts in \cite{clements10}. Likewise, the LoBAL number counts appear consistent with models from both \cite{lagache04} and \cite{le09}. Based on Fig.~\ref{fig:nCounts} we therefore find no evidence for an overdensity in the environments of our LoBAL sample on scales of $\sim$1Mpc, concluding instead that LoBALs reside in environments consistent with the general galaxy population at 2.0$<$z$<$2.5. We highlight that although the lack of enhancement in the environment of LoBALs is consistent with the triggering of LoBALs via secular processes, such as bar instabilities, minor mergers and stochastic gas accretion, this observation does not necessarily rule out the triggering of these quasars via gas-rich major mergers. Testing the environments of LoBALs, as we have done here, is an indirect test for the presence of mergers. Indeed, several studies of the highest-luminosity quasars, most often associated with merger triggering, do not find these systems to lie in the highest environmental overdensities \citep{fanidakis13, uchiyama18}. Future, high-resolution imaging of these systems is required to directly trace merger signatures (e.g. morphological disruptions) in our LoBAL sample. Furthermore, the resolution of the Herschel images used throughout this work means we may be susceptible to source blending in some cases (see Sec.~\ref{sec:caveats}), resulting in the environmental density being underestimated.

\subsection{Caveats}
\label{sec:caveats}

Here, we discuss the potential caveats in our investigation. First and foremost, the effects of source blending have not been considered in our analysis, meaning we cannot rule out the possibility that the measured fluxes arise from multiple sources in the image, particularly at 500$\mu$m where the resolution drops to 14 arcsec pix$^{-1}$. Although we emphasise that this potential source blending does not affect the direct comparisons of the detection rates to the HiBAL and non-BAL samples (Sec.~\ref{sec:CompSamples}), we note that it will likely result in the overestimation of the SFRs in our sample. As such, we highlight that the the SFRs derived in this paper are likely upper estimates on the true values. However, ALMA observations of IR-bright quasars imaged with Herschel SPIRE \citep{hatziminaoglou18} reveal that the majority of FIR-bright quasars are not affected by source blending. Of the 28 quasars in their sample, just 30 per cent consist of multiple sources, with the companion galaxy contributing to the flux at 870$\mu$m. The remaining 70 per cent appear to uniquely lie within the SPIRE beam, meaning that source blending is unlikely to be an issue for the majority of IR-bright quasars such as those in our sample. Furthermore, a recent study by \cite{liu18} developed an algorithm aimed at addressing any potential source blending by accurately measuring multi-wavelength photometry from highly confused images. Although this currently exists over a limited field and thus cannot be applied to our LoBAL sample, we note that future studies may benefit from such tools.

Secondly, whilst our LoBAL sample overlaps the comparison sample of HiBAL quasars \citep{cao12} in magnitude and redshift, the two samples are not fully matched, with the LoBAL targets generally lying towards the bright, high-redshift end of the HiBAL sample. Furthermore, the LoBAL $i$-band magnitudes plotted in Fig.~\ref{fig:HiBALcompare} have not been corrected for dust attenuation, meaning the difference in luminosity may be even more pronounced. It is likely therefore, that the enhancement in the detection rate of LoBALs we measure is strictly an upper limit. We also note that the SFR, and thus the FIR flux, strongly evolves with redshift, increasing out to a redshift of z$\sim$2 \citep[e.g.][]{madau14}. This redshift bias may therefore partially account for the enhancement we observe in the detection rates of LoBALs with regards to HiBAL quasars. We highlight however, that both the HiBAL and LoBAL samples lie at the peak of cosmic star formation at z$\sim$1-3 \citep{madau14}, and thus the redshift evolution of the SFR will be minimal at this epoch. Whilst the reshift bias is therefore unlikely to fully account for the detection rate enhancement we observe, we suggest that the enhancement factor of 8.5 shown in Fig.~\ref{fig:loHi} is likely an overestimate.

\section{Conclusions}
\label{sec:Conclusions}

We have presented targeted Herschel SPIRE observations for a sample of 12 LoBALs at 2.0$<$z$<$2.5. Our key conclusions are as follows;

(i) Three of the 12 LoBAL targets (25 per cent) are detected with \textit{Herschel} at $>$5$\sigma$ in all the SPIRE bands (250, 350 and 500$\mu$m). Based on the simultaneous quasar + galaxy SED fitting of the combined WISE + PACS + SPIRE photometry for these targets (Fig.~\ref{fig:SEDs}), we infer high SFRs ranging 740 - 2380M$_{\odot}$yr$^{-1}$ and indicating strong star formation activity in LoBALs at 2.0$<$z$<$2.5. Furthermore, stacking the non-detections and assuming FIR emission to be dominated by cold dust places a 3$\sigma$ upper limit on the SFR of $<$440M$_{\odot}$yr$^{-1}$, meaning even among the undetected sources we cannot rule out moderate-to-high SFRs in LoBALs.

(ii) We find evidence for an enhancement in the FIR detection rate of LoBALs compared to populations of both HiBAL and non-BAL quasars. When considering sources detected in all \textit{Herschel} SPIRE bands, we derive enhancement factors  $\sim$8.5 and $\sim$1.5 with regards to the HiBAL and non-BAL samples respectively. This result indicates a likely enhancement of star formation in LoBALs compared to other quasar populations, supporting an evolutionary picture of LoBALs in which they exist in a short-lived phase following starburst activity in the galaxy. Indeed, direct comparisons between the SFRs in non-BALs with our LoBAL sample reveal such an enhancement in the SFRs of LoBALs.

(iii) Despite detecting several serendipitous sources within 1.5arcmin of the LoBAL targets, we find no statistical differences in the local environments of LoBALs compared to the H-ATLAS blank fields in any of the SPIRE bands. Thus, we find no evidence to suggest that LoBALs exist in any `special' FIR environment at 2.0$<$z$<$2.5, rather they appear to reside in environments typical of the general galaxy population at this redshift. This lack of distinction in the environments of LoBALs potentially supports an orientation model of LoBALs, in which their local environments are expected to be consistent with other quasar populations.

Overall, we find tentative evidence that LoBALs exist in a special evolutionary phase, although we cannot rule out an orientation scenario. Future high-resolution imaging (e.g. from ALMA) at $\lambda$ $>$ 500$\mu$m will enable emission from the galaxy to be isolated and thus improve our understanding of star formation in LoBALs at 2.0$<$z$<$2.5 and thus their role in quasar evolution.

\section*{Acknowledgements}
CFW and JK acknowledge financial support from the Academy of Finland, grant 311438. 
Herschel is an ESA space observatory with science instruments provided by European-led Principal Investigator consortia and with important participation from NASA.
SPIRE has been developed by a consortium of institutes led
by Cardiff Univ. (UK) and including Univ. Lethbridge (Canada); NAOC (China);
CEA, LAM (France); IFSI, Univ. Padua (Italy); IAC (Spain); Stockholm
Observatory (Sweden); Imperial College London, RAL, UCL-MSSL, UKATC,
Univ. Sussex (UK); Caltech, JPL, NHSC, Univ. Colorado (USA). This develop-
ment has been supported by national funding agencies: CSA (Canada); NAOC
(China); CEA, CNES, CNRS (France); ASI (Italy); MCINN (Spain); SNSB
(Sweden); STFC (UK); and NASA (USA).
This publication also makes use of data products from the Wide-field Infrared Survey Explorer, which is a joint project of the University of California, Los Angeles, and the Jet Propulsion Laboratory/California Institute of Technology, funded by the National Aeronautics and Space Administration. Black hole masses are based on observations made with the Nordic Optical Telescope, operated by the Nordic Optical Telescope Scientific Association at the Observatorio del Roque de los Muchachos, La Palma, Spain, of the Instituto de Astrofisica de Canarias.

\section*{Data Availability}
The data underlying this article are available in the ESA Herschel Science Archive, at http://archives.esac.esa.int/hsa.




\bibliographystyle{mnras}
\bibliography{refNew} 







\bsp	
\label{lastpage}
\end{document}